\documentclass{aa}  

\usepackage{newtxtext,newtxmath}

\usepackage[T1]{fontenc}
\usepackage{ae,aecompl}

\usepackage{multirow}
\usepackage{adjustbox}

\usepackage[verbose]{placeins}
\usepackage{blindtext}

\usepackage{bm}
\usepackage{mathtools}
\usepackage{xcolor}
\usepackage{natbib}
\usepackage{lscape}
\usepackage{afterpage}  
\usepackage{booktabs}
\usepackage{tabularx}
\usepackage{threeparttable}

\usepackage{newtxtext,newtxmath}
\usepackage{graphicx}

\begin{document} 

   \title{Witnessing an extreme, highly efficient galaxy formation mode with resolved Ly$\alpha$ and LyC emission }
    \titlerunning{High-resolution imaging of J1316+2614}
   
\author{R.~Marques-Chaves\inst{\ref{inst1}} \thanks{e-mail: Rui.MarquesCoelhoChaves@unige.ch}
    \and
    D.~Schaerer\inst{\ref{inst1},\ref{inst2}}
    \and
    E.~Vanzella\inst{\ref{inst3}}
    \and
    A.~Verhamme\inst{\ref{inst1}}
    \and
    M.~Dessauges-Zavadsky\inst{\ref{inst1}}
    \and
    J.~Chisholm\inst{\ref{inst4}}
    \and
    F.~Leclercq\inst{\ref{inst4}}
    \and
    A.~Upadhyaya\inst{\ref{inst5}}
    \and
    J.~Álvarez-Márquez\inst{\ref{inst6}}
    \and
    L.~Colina\inst{\ref{inst6}}
    \and
    T.~Garel\inst{\ref{inst1}}
    \and
    M.~Messa\inst{\ref{inst3}}
}

\institute{
    Department of Astronomy, University of Geneva, 51 Chemin Pegasi, 1290 Versoix, Switzerland \label{inst1}
    \and
    CNRS, IRAP, 14 Avenue E. Belin, 31400 Toulouse, France \label{inst2}
    \and
    INAF – OAS, Osservatorio di Astrofisica e Scienza dello Spazio di Bologna, via Gobetti 93/3, I-40129 Bologna, Italy \label{inst3}
    \and
    Department of Astronomy, University of Texas at Austin, 2515 Speedway, Austin, Texas 78712, USA \label{inst4}
    \and
    Department of Physics, University of Warwick, Gibbet Hill Road, Coventry CV4 7AL, UK \label{inst5}
    \and
    Centro de Astrobiología (CAB), CSIC-INTA, Ctra. de Ajalvir km 4, Torrejón de Ardoz, E-28850, Madrid, Spain  \label{inst6}    
}

   \date{Received --; accepted --}
 
   \abstract

\abstract{
J1316+2614 at $z$=3.613 is the UV-brightest ($M_{\rm UV} = -24.7$) and strongest Lyman continuum (LyC, $f_{\rm esc}^{\rm LyC} \approx $90\%) emitting star-forming galaxy known, showing also signatures of inflowing gas from its blue-dominated Ly$\alpha$ profile. Here, we present high-resolution imaging with the \textit{Hubble} Space Telescope (\textit{HST}) and the Very Large Telescope (VLT) of the LyC, Ly$\alpha$, rest-UV, and optical emission of J1316+2614. 
Detailed analysis of the LyC and UV light distributions reveals compact yet resolved profiles, with LyC and UV morphologies showing identical half-light radii of $r_{\rm eff} \simeq 220$\,pc. 
The continuum-subtracted Ly$\alpha$ emission, obtained with the \textit{HST} ramp-filter FR551N, reveals an extended filamentary structure of $\simeq 6.0$\,kpc oriented south-north with only weak/residual flux within the stellar core, suggesting a Ly$\alpha$ "hole". Our SED analysis shows that J1316+2614 is characterized by a young (5.7$\pm$1.0\,Myr), nearly un-obscured stellar population with a high star formation rate ($\rm SFR = 898 \pm 181$ $M_{\odot}$\,yr$^{-1}$) and a stellar mass of $M_{\star}^{\rm young} = (4.8 \pm 0.3) \times 10^{9} M_{\odot}$. Additionally, the SED analysis supports the absence of an underlying old stellar population ($M_{\star}^{\rm old} \leq 2.8 \times 10^{9} M_{\odot}$, 3$\sigma$). J1316+2614 presents remarkably high SFR and stellar mass surface densities of log($\Sigma SFR [M_{\odot} \rm yr^{-1} kpc^{-2}]) = 3.47 \pm 0.11$ and log($\Sigma M_{\star} [M_{\odot} \rm pc^{-2}]) = 4.20 \pm 0.06$, respectively, which are among the highest observed in star-forming galaxies and are more typically observed in local young massive star clusters and Globular clusters.
Our findings indicate that J1316+2614 is a powerful, young, and compact starburst, leaking significant LyC photons due to the lack of gas and dust within the starburst. 
We explore the conditions for gas expulsion using a simple energetic balance and find that, given the strong binding force in J1316+2614, a high star formation efficiency ($\epsilon_{\rm SF} \geq 0.7$) is necessary to remove the gas and explain its exposed nature. Our results thus suggest a close link between high $\epsilon_{\rm SF}$ and high $f_{\rm esc}^{\rm LyC}$. This high efficiency can also naturally explain the remarkably high SFR, UV-luminosity, and efficient mass growth of J1316+2614, where at least 62\% of its mass formed in the last 6\,Myr. J1316+2614 may exemplify an intense, feedback-free starburst with a high $\epsilon_{\rm SF}$, similar to those proposed for UV-bright galaxies at high redshifts. 
}

    \maketitle
%

\section{Introduction}

UV-bright star-forming galaxies were once considered extremely rare at any redshift, even at the Epoch of Reionization (EoR, at $6 < z < 16$). However, recent \textit{James Webb} Space Telescope (\textit{JWST}) observations have dramatically changed this picture by revealing a large number of UV-bright and sometimes massive galaxies at $z \simeq 7 - 14$ \citep[see e.g.,][for some spectroscopically confirmed sources]{ArrabalHaro2023, Bunker2023, Carniani2024, Castellano2024}. The derived volume densities of these sources exceed predictions from galaxy formation models and pre-\textit{JWST} observations by an order of magnitude \citep[e.g.,][]{Bouwens2021, Kannan2023, Lovell2023MNRAS}. These results are enigmatic based on the common wisdom of galaxy formation and evolution, challenging our understanding of the nature of UV-bright galaxies and the potential role these sources play in cosmic reionization.

Several scenarios have been proposed to explain this tension. One suggestion is that the star formation efficiency ($\epsilon_{\rm SF}$), i.e., the efficiency in converting gas into stars, is higher than assumed in current models and measured locally (of a few percent, e.g., \citealt{Megeath2016}). In this framework \citep{Dekel2023, Li2023, Boylan-Kolchin2024}, high-density environments and low metallicities, properties expected at early times, may favor the formation of so-called “feedback-free starbursts” \citep{Dekel2023} through the collapse of gas clouds within very short free-fall times. This would increase the star formation efficiency since the cloud collapse occurs before the onset of strong feedback, thus enhancing star-formation rates (SFRs), UV luminosities, and stellar masses. Other works have related the excess of UV-bright sources to variations of the initial mass function (IMF) that allow the formation of more massive stars \citep[e.g.,][]{ Inayoshi2022, Finkelstein2023, Trinca2024}. This excess of massive stars, also referred to as a “top-heavy” IMF, boosts the UV radiation and the luminosity-to-mass ratio making these sources appear (UV) brighter at fixed mass. On the other hand, \cite{Ferrara2023} (see also \citealt{Ziparo2023}) proposed that radiation-driven outflows originating from recent star formation could temporarily remove dust as soon as it is produced. Dust ejection by strong radiative feedback would decrease the dust optical depth, making these galaxies appear brighter in the UV. Other frameworks invoke the stochastic nature of star formation at high redshifts \citep[e.g.,][]{Mason2023, Shen2023}, or even the contribution from active galactic nuclei \citep{Hegde2024, Maiolino2024}. 

In parallel, extremely UV-bright star-forming galaxies at $z \simeq 2-4$ have been discovered in the wide Sloan Digital Sky Survey by \cite{marques2020b, marques2021, marques2022}. These galaxies present remarkably high UV absolute magnitudes of $M_{\rm UV} \sim -24$ and are characterized by very young ($\leq 10$\,Myr) stellar populations without signs of AGN activity by the detection of photospheric absorption lines and wind line features and UV/optical BPTs. They show star-formation rates up to $\rm SFR \simeq 1000$\,$M_{\odot}$\,yr$^{-1}$, but residual dust attenuation with UV continuum slopes as steep as $\beta_{\rm UV} \simeq -2.6$ \citep[e.g.,][]{marques2022}. As such, these sources are among the most vigorous and almost un-obscured star-forming galaxies known, with specific star formation rates sSFR $>50-100$\,Gyr$^{-1}$. Furthermore, they also show complex gas kinematics, including outflows \citep{alvarez2021, marques2021} and inflows \citep{marques2022}. Recently, the analysis of the rest-UV spectra of these sources by \cite{Upadhyaya2024} have revealed signatures of very massive stars (VMS, with initial masses $>100 M_{\odot}$) for most of them, suggesting that VMS might be common in UV-bright galaxies. Last but not least, the few sources with Lyman continuum (LyC) observations so far, J0121+0025 ($z=3.2$) and J1316+2614 ($z=3.6$), showing copious LyC leakage, with absolute escape fractions up to $f_{\rm esc}^{\rm LyC} \approx 90\%$ \citep{marques2021,marques2022}. As the UV-brightest star-forming galaxies known, they are ideal laboratories to test the various scenarios proposed to explain the overabundance of UV-bright EoR sources.

Here we present high spatial-resolution observations of J1316+2614 at $z=3.61$ \citep{marques2022}, the UV-brightest star-forming galaxy ($M_{\rm UV}=-24.7$) and also the strongest LyC emitter known ($f_{\rm esc}^{\rm LyC} \simeq 90\%$). 
J1316+2614 is a powerful starburst with negligible dust attenuation given by its steep UV slope ($\beta_{\rm UV} = -2.59 \pm 0.05$). It shows relatively weak nebular emission (e.g., $EW_{0} (\rm H\beta) = 34.7 \pm 6.8$\,\AA) due to the high fraction of ionizing photons escaping its interstellar medium ($f_{\rm esc}^{\rm LyC} \approx 90\%$ and log($Q_{\rm H}^{\rm esc} / s^{-1}) = 55.86 \pm 0.11$, \citealt{marques2022}). Finally, J1316+2614 also shows a peculiar Ly$\alpha$ spectral profile with a blue-to-red peak line ratio $I_{\rm blue}/I_{\rm red} \simeq 3.7$, suggesting inflows.

This work is organized as follows. In Section \ref{observations} we described high-resolution observations taken with the Hubble Space Telescope (\textit{HST}) and the Very Large Telescope (VLT) probing the LyC, Ly$\alpha$, and the rest-UV and optical emission of J1316+2614. In Section \ref{results} we describe the methodology and present the main results, including the morphology, photometry, and the spectral energy distribution (SED) of J1316+2614. The discussion of the results is presented in Section \ref{discussion}, and, finally, we present the summary of our main findings in Section \ref{conclusion}. Throughout this work, we use a concordance cosmology with $\Omega_{\rm m} = 0.274$, $\Omega_{\Lambda} = 0.726$, and $H_{0} = 70$ km s$^{-1}$ Mpc$^{-1}$. Magnitudes are given in the AB system.

\section{Observations}\label{observations}

\subsection{\textit{HST} imaging}

High-resolution imaging of J1316+2614 was obtained with the UVIS/IR imager Wide Field Camera 3 (WFC3) and the Advanced Camera for Surveys (ACS) of Wide Field Channel (WFC) aboard the \textit{HST}. These observations were carried out between June 26 and July 3 2023 under the Cycle 30 program ID 17286 (PI: R. Marques-Chaves). J1316+2614 was observed with the WFC3 in the medium-band F410M and broad-band filters F775W and F160W, with total exposure times of 5004\,s, 2372\,s, and 2412\,s, respectively. These filters probe the rest-frame LyC ($\simeq$871-914\,\AA)\footnote{The F410M filter response at $\lambda_{0}>912$\,\AA{ }is less than $\sim 0.08$, and therefore, the contamination of non-LyC emission in F410M is negligible (e.g., \citealt{Smith2018}).}, UV ($\simeq$1650\,\AA), and optical ($\simeq$3310\,\AA) emission of J1316+2614. Additional ACS/WFC observations were obtained with the narrow-band ramp filter FR551N (transmission width of $\simeq 97$\,\AA) centered at $\lambda_{\rm obs} = 5604$\,\AA{ }to cover the Ly$\alpha$ emission of J1316+2614 at $z=3.612$. The ACS total exposure time was 2008\,s. Table \ref{tab1} summarizes the \textit{HST} observations.

\begin{figure}
  \centering
  \includegraphics[width=0.45\textwidth]{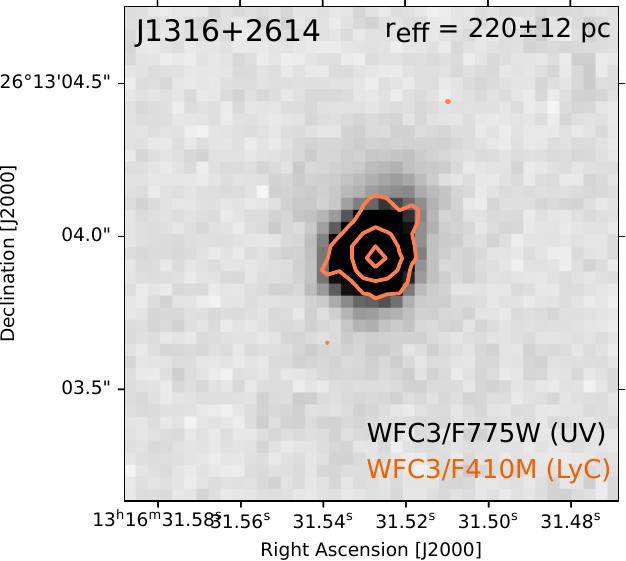}
  \caption{Cutout image of J1316+2614 showing the \textit{HST} F775W (background image) and F410M (orange contours, 5$\sigma$, 10$\sigma$, and 50$\sigma$) which probe the rest-UV and LyC emission of J1316+2614, respectively. }
  \label{fig1}
\end{figure}

Data were reduced using \texttt{AstroDrizzle} version 3.6.2 from the \texttt{DrizzlePac} package \citep{Fruchter2002} and retrieved from MAST. The final images have pixel scales of $0.04^{\prime \prime}$ pix$^{-1}$ (F410M and F775W), $0.05^{\prime \prime}$ pix$^{-1}$ (FR551N), and $0.12^{\prime \prime}$ pix$^{-1}$ (F160W). The astrometry was corrected and aligned to the GAIA DR3 \citep{Gaia2023}. The astrometry r.m.s precision is $\simeq 0.09^{\prime \prime}- 0.12^{\prime \prime}$. The instrumental point-spread function (PSF) in each image was modeled using the Point Spread Function reconstruction (\texttt{PSFr}, \citealt{Birrer2021, Birrer2022}) code by stacking several (3-5) isolated bright stars within the field of view (FoV) of the observations. We measured PSF FWHM of 0.075$^{\prime \prime}$, 0.109$^{\prime \prime}$, 0.080$^{\prime \prime}$, 0.197$^{\prime \prime}$ for F410M, FR551N, F775W, and F160W, respectively. Figure \ref{fig1} shows the F775W image of J1316+2614 and the contours from F410M.

\begin{table*}
\begin{center}
\caption{Summary of the \textit{HST} and VLT imaging observations of J1316$+$2614. \label{tab1}}
\resizebox{1.0\textwidth}{!}{%
\begin{tabular}{l c c c c c c c c c}
\hline \hline
\smallskip
Filter & $\lambda_{\rm rest}$ [\AA] &  $t_{\rm exp}$ [s] & Scale [$^{\prime \prime}$/pix] & PSF$_{\rm FWHM}$ [$^{\prime \prime}$] & Mag. [AB] & $n$ & $b/a$ & $\theta$ [deg] & $r_{\rm eff}$ [pc] \\
$\rm (1)$ & (2) & (3) & (4) & (5) & (6) & (7) & (8) & (9) & (10) \\
\hline 
F410M & 871-914 & 5004 & 0.0394 & 0.0745 & $23.32 \pm 0.06$ & $2.12 \pm 0.91$ & $0.71 \pm 0.19$ & $198 \pm 36$ & $262 \pm 64$ \\
FR551N & 1204-1225 & 2008 & 0.0500 & 0.1085 &  $20.81 \pm 0.10$  & --- & --- & --- & --- \\
F775W & 1489-1859 & 2372 & 0.0396 & 0.0796 & $21.25 \pm 0.04$ & $2.90 \pm 0.42$ &  $0.72 \pm 0.05$ & $162 \pm 5$ & $220 \pm 12$ \\
F160W & 3004-3686 & 2412 & 0.1272 & 0.1968 &  $21.66 \pm 0.05$  & --- & --- & --- & $\leq 442$ \\
$K_{\rm s}$ & 4301-5003 & 1350 & 0.1065 & 0.2960 & $21.73 \pm 0.06$ & --- & --- & --- & $\leq 550$ \\

\hline 
\end{tabular}%
}
\end{center}
\textbf{Notes. ---} (1) and (2) Filter and corresponding bandwidth in the rest-frame; (3) exposure time; (4) pixel scale; (5) PSF FWHM obtained from stars in the field-of-view; (6) aperture photometry of J1316+2614; (7) Sersic index; (8) minor-to-major axis; (9); orientation (north=0, east=90); and (10) half-light effective radius. 
\end{table*}

\subsection{VLT/HAWK-I imaging}

Additional near-IR imaging of J1316+2614 was obtained in the $K_{\rm s}$-band with the HAWK-I on the VLT UT4. These observations were conducted on May 31 2023 as part of the program ID 111.251K.001 (PI: R. Marques-Chaves). $K_{\rm s}$-band observations were obtained with the GRound layer Adaptive optics system Assisted by Lasers (GRAAL), enhancing the final image quality down $\simeq 0.296^{\prime \prime}$, as measured from the light profiles of several stars in the HAWK-I FoV. The on-source exposure time was 1350\,s. Data were reduced using the standard ESO pipeline version 2.4.12 \footnote{\url{https://www.eso.org/sci/software/pipelines/hawki/hawki-pipe-recipes.html}.} and were flux calibrated against 2MASS stars in the field. The astrometry was calibrated using the GAIA DR3 catalog \citep{Gaia2023} yielding an r.m.s precision of $\simeq 0.10^{\prime \prime}$, similar to the native pixel-scale (0.107$^{\prime \prime}$).

\section{Methodology and Results}\label{results}

\begin{figure*}
  \centering
  \includegraphics[width=1\textwidth]{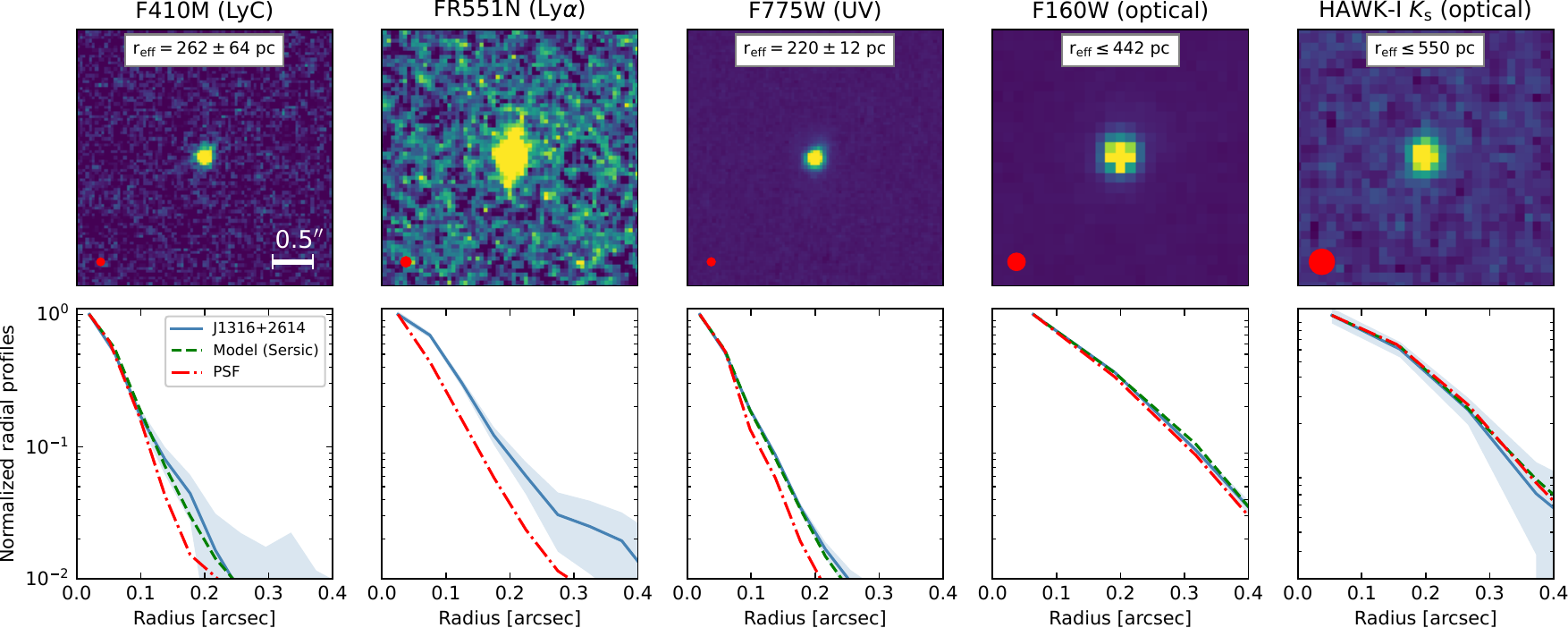}
  \caption{\textit{HST} and VLT images of J1316+2614. From left to right, we show the \textit{HST} F410M (LyC), FR551N (Ly$\alpha$), F775W (rest-UV), F160W (rest-optical), and VLT HAWK-I $K_{\rm s}$ (rest-optical) images. The FWHM PSF of image is represented with a red circle. Each stamp has a size of $3.2^{\prime \prime} \times 3.2^{\prime \prime}$. North is up and East is to the left. The bottom panels show the normalized radial profiles (to their maxima) of J1316+2614 in each band (solid blue with uncertainties in shadow), the best-fit Sersic model (dashed green), and the PSF used in the fit (red dotted dashed). J1316+2614 shows a compact stellar morphology, only resolved in F410M and F775W ($r_{\rm eff} \simeq 220$\,pc). The FR551N, which probes the Ly$\alpha$ emission (and stellar continuum), shows a more extended morphology.}
  \label{fig2}
\end{figure*}

\subsection{Size measurements}\label{morphology}

As shown in Figure \ref{fig2}, the \textit{HST} and VLT images reveal a compact morphology in the bands probing the stellar continuum of J1316+2614 (F410M, F775W, F160W, and $K_{\rm s}$). In contrast, the ACS/FR551N image, which predominantly traces the Ly$\alpha$ emission, shows a well-resolved and extended profile. 

\subsubsection{Stellar morphology}\label{lyc_morphology}

The light distribution of J1316+2614 is investigated using the \texttt{PySersic} code \citep{Pasha2023}, which uses a Bayesian framework to understand the degeneracies between different parameters. \texttt{PySersic} fits the light distribution of a source using morphological models convolved with a given PSF. As described in Section \ref{observations}, the PSF of each image/band is obtained by stacking several bright stars within the FoV using the \texttt{PSFr} code \citep{Birrer2021, Birrer2022}. We fit the morphology of J1316+2614 using both 2D Sersic (with a Sersic index varying from 0.5 to 6.0) and point-like profiles to investigate whether the light distribution of J1316+2614 is resolved in each band. 

\begin{figure}
  \centering
  \includegraphics[width=0.48\textwidth]{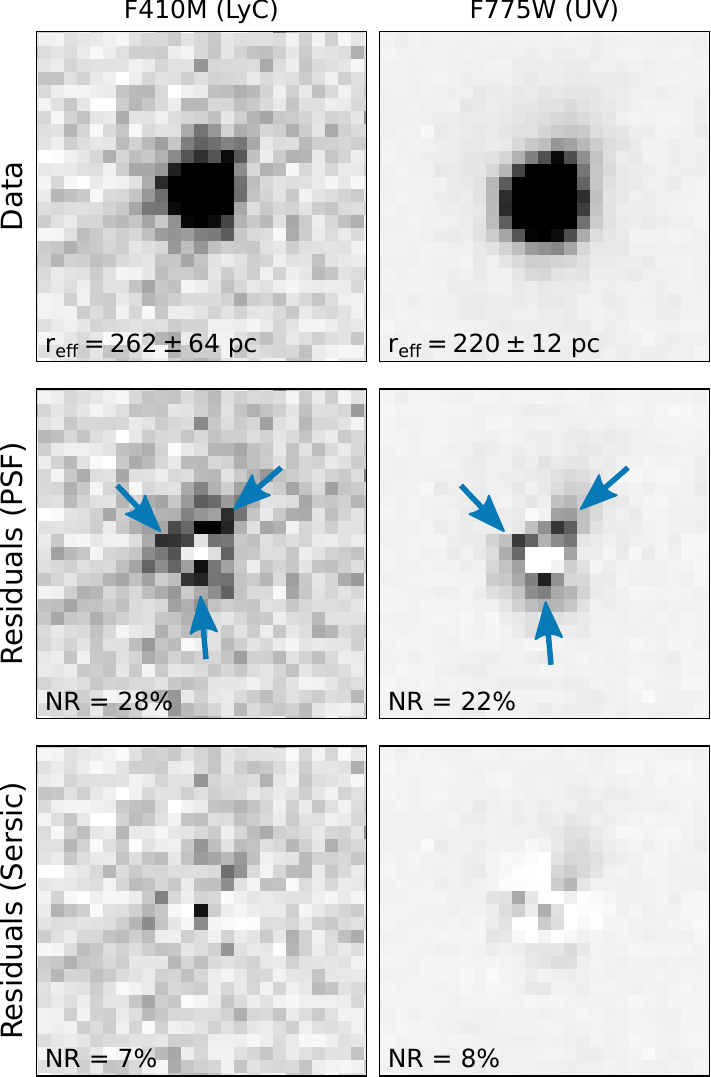}
  \caption{\textit{HST} F410M (LyC, left) and F775W (UV, right) images of J1316+2614 (top) and the residuals obtained after subtracting the PSF and Sersic best-fit models (middle and bottom, respectively). The normalized residuals (NR), measured around a circular aperture of 0.6$^{\prime \prime}$ around J1316+2614, are also indicated. Each stamp has a size of $1.0^{\prime \prime} \times 1.0^{\prime \prime}$. North is up and East is to the left. The PSF-subtracted residuals are identical in F410M and F775W, indicating similar LyC and rest-UV morphologies.}
  \label{fig2_new}
\end{figure}

We start fitting the light profile of J1316+2614 using the \textit{HST} F775W (rest-UV). Assuming a Sersic profile, \texttt{PySersic} finds an effective radius $r_{\rm eff} = 0.76 \pm 0.04$\,pix (or $r_{\rm eff} = 220 \pm 12$\,pc with our adopted cosmology) and a Sersic index $n=2.90 \pm 0.42$. The normalized residuals (NR), measured within a circular aperture of 0.6$^{\prime \prime}$ around J1316+2614, are $\rm NR \simeq 8\%$ (Fig. \ref{fig2_new}). While a Sersic profile recovers most ($\gtrsim 90\%$) of the light emission in F775W, the model-subtracted image shows some residuals that are not perfectly accounted for and may suggest additional underlying structures. 
If instead a point-like profile is used in the fit, \texttt{PySersic} cannot recover well the light profile of J1316+2614 yielding substantial residuals in the model-subtracted image ($\rm NR \simeq 22\%$, right middle panel of Figure \ref{fig2_new}). Our results thus indicate that J1316+2614 has a resolved morphology in the rest-UV continuum ($r_{\rm eff} = 220 \pm 12$\,pc), as also indicated through its radial profile (bottom panel of Fig. \ref{fig2}).

Similar to F775W, the profile of J1316+2614 in the \textit{HST} F410M (LyC) appears resolved. Assuming a Sersic profile, the \texttt{PySersic} best-fit predicts an effective radius $r_{\rm eff} = 0.79 \pm 0.21$\,pix or $r_{\rm eff} = 262 \pm 64$\,pc and a Sersic index $n=2.12 \pm 0.91$. The normalized residuals from this fit, measured within $r\leq 0.6^{\prime \prime}$), are significantly better ($\simeq 7\%$) than those obtained assuming a point-like source ($\simeq 28\%$, Fig. \ref{fig2_new}). This suggests that the LyC emission seen in F410M is resolved and has a similar morphology to the rest-UV emission. To investigate this, we inspect the residuals obtained from the PSF-subtracted images. As shown in Figure \ref{fig2_new}, the PSF-subtracted images in F410M and F775W show almost identical residuals (highlighted with blue arrows), indicating that the even faintest resolved emission in F775W is present in F410M. In addition, we inspect the F410M and F775W normalized radial profiles of J1316+2614 and find that they are indistinguishable within the uncertainties (Figure \ref{fig2}). Finally, we model the F410M emission using the F775W image of J1316+2614 as a PSF and assuming a point-like source. The normalized residuals are slightly better than the ones obtained assuming a Sersic model. Altogether, our results indicate that the LyC (F410M) and rest-UV (F775W) light profiles of J1316+2614 are essentially the same, and consistent with $r_{\rm eff} = 220$\,pc. This strongly supports that the LyC and UV emissions have similar origins. 

Turning to longer wavelengths, the light distribution of J1316+2614 in the \textit{HST} F160W and VLT $K_{\rm s}$ bands appears unresolved. This is expected given the slightly poorer spatial resolution in these bands and the fact that they still probe the young stellar population of J1316+2614 ($r_{\rm eff} \simeq 220$\,pc). 
Our best-fit models assuming Sersic or point-like profiles yield essentially similar residual images. Given the oversampling of the \textit{HST}/F160W PSF ($\rm FWHM \simeq 1.6$\,pix) we use the minimum resolvable size of $r_{\rm eff} \leq 0.47$\,pix estimated in \cite{Messa2022} for the same instrument and filter considered here to infer the upper limit $r_{\rm eff} \leq 442$\,pc in the \textit{HST}/F160W. Since the HAWK-I PSF is well sampled ($\rm FWHM \simeq 2.8$ pix), we derive the upper limit in the $K_{\rm s}$-band of $r_{\rm eff} \leq 550$\,pc assuming $\rm FWHM_{\rm min} \leq \rm FWHM (PSF) / 2$.

\subsubsection{Ly$\alpha$ morphology}

\begin{figure*}
  \centering
  \includegraphics[width=0.73\textwidth]{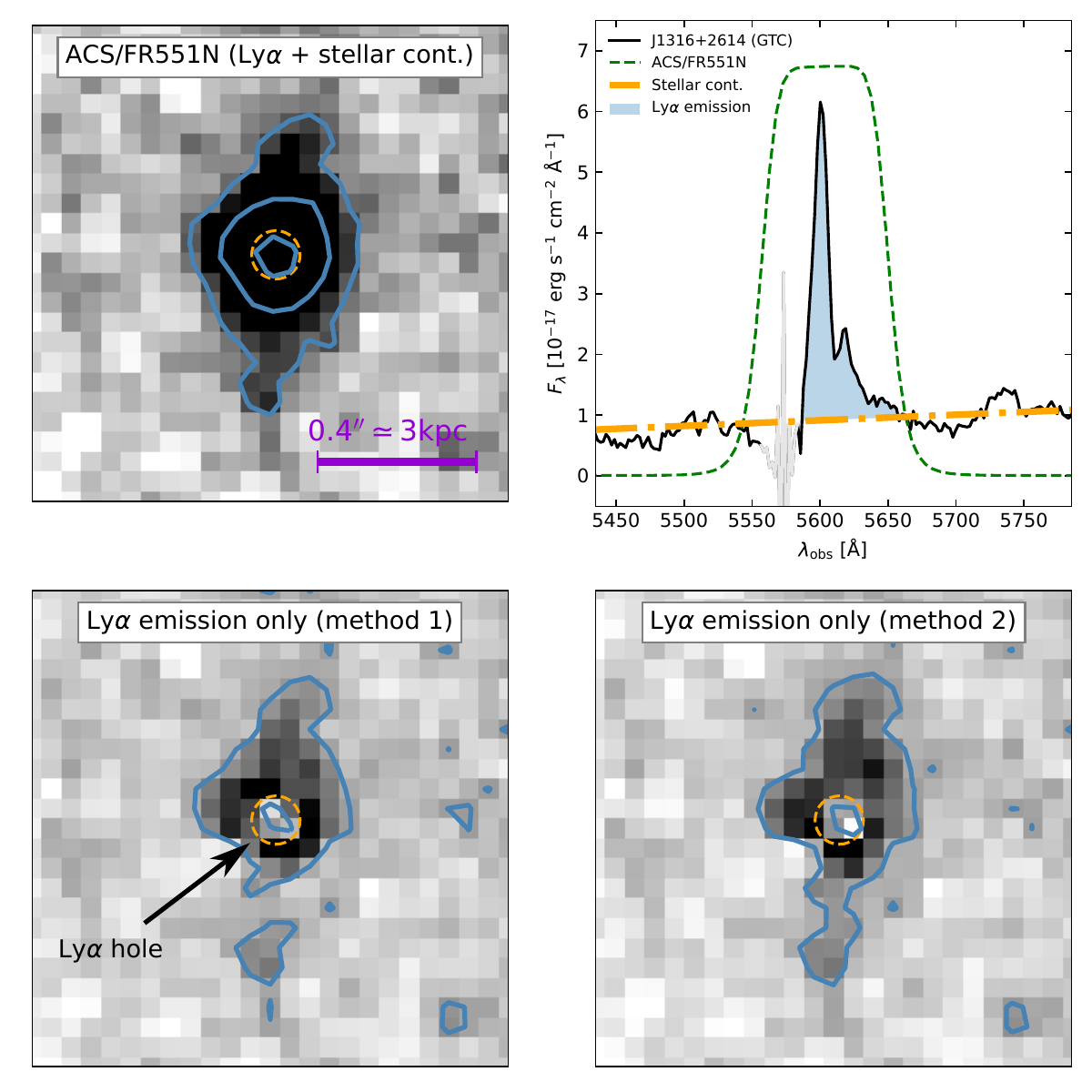}
  \caption{Ly$\alpha$ spatial distribution of J1316+2614. The top left panel shows the cutout images of J1316+2614 in the ACS/\textit{HST} ramp-filter FR551N (with a total size of $1.2^{\prime \prime} \times 1.2^{\prime \prime}$, North is up and East to the left), which includes the Ly$\alpha$ emission and the underlying stellar continuum (blue contours mark the 3$\sigma$, 15$\sigma$, and 50$\sigma$ emission). The orange dashed circle represents the position and total size of the stellar continuum as measured in the F775W after deconvolved with the PSF (i.e., radius of $2\times r_{\rm eff} \simeq 0.06^{\prime \prime}$). The top right panel shows the GTC spectrum of J1316+2614 \citep[black,][]{marques2022} and the FR551N transmission curve (dashed green). The orange dashed-dotted line represents our best fit of the stellar continuum around the Ly$\alpha$ emission (blue). The bottom panels show the continuum-subtracted FR551N images using two different methods (see text). Ly$\alpha$ appears weak/residual within the UV-bright stellar clump (orange) and is predominantly emitted in the outskirts. The blue contours mark the 2.5$\sigma$ level.}
  \label{fig3}
\end{figure*}

Finally, we analyze the \textit{HST} ACS/FR551N image of J1316+2614 which includes the Ly$\alpha$ emission ($EW_{0} = 20.5 \pm 1.9$\,\AA, \citealt{marques2022}). As shown in the top left panel of Figure \ref{fig3}, the ACS/FR551N image reveals a complex morphology consisting of a bright central clump co-spatial with the compact stellar emission (e.g., as seen in F410M or F775W) and a more diffuse, filamentary-like emission oriented south-north with a $3\sigma$ scale length of $\simeq 0.8^{\prime \prime}$ or $\simeq 6.0$ kpc. 

We employ two different methodologies to subtract the underlying stellar emission in the FR551N band. For the first one (method 1 in Figure \ref{fig3}), we use the WFC3/F775W as the reference image of the stellar emission of J1316+2614, which is first resampled to the FR551N native pixel size (0.05$^{\prime \prime}$\,pix$^{-1}$) using the \texttt{MAGNIFY} task from \texttt{Iraf}. We repeat this step in an individual star and find no significant differences between the PSF FWHM measured in FR551N and the resampled F775W images.
Since the astrometry uncertainties in both filters ($\simeq 0.10^{\prime \prime}$) are larger than the pixel size, we choose to spatially match both images of J1316+2614 using their centroid emission and the corresponding shifts in pixels using the \texttt{Iraf} task \texttt{imshift}. This step assumes that the centroid emission in FR551N is dominated by the stellar continuum, which is a fair assumption since the continuum emission represents $\simeq 50\%$ of the total flux in FR551N (see next) and is way more compact than the extended Ly$\alpha$ emission. 
We infer the contribution of the stellar continuum in the FR551N passband using the low-resolution optical spectrum of J1316+2614 obtained with GTC/OSIRIS, which was previously rescaled to the $R$-band photometry to account for slit losses \citep[see][]{marques2022}. As highlighted in the top right panel of Figure \ref{fig3}, the contribution of the stellar emission is obtained by fitting a linear polynomial function using two spectral windows on each side of Ly$\alpha$ (5480-5527\,\AA{ }and 5757-5841\,\AA). 
Using \texttt{PyPhot}\footnote{\url{https://github.com/mfouesneau/pyphot}} and the FR551N transmission profile centered at $\lambda = 5604$\,\AA, we measure the synthetic photometry of the polynomial function, which probes only the stellar continuum. We find $f_{\nu}^{\rm cont} \simeq 9.62 \times 10^{-29}$ erg\,s$^{-1}$\,cm$^{-2}$\,Hz$^{-1}$ which represents roughly $\simeq 50\%$ of the total emission in FR551N. Finally, we rescale the flux of the resampled F775W image to that obtained from the synthetic photometry and subtract it from the FR551N image. 

The bottom left panel of Figure \ref{fig3} shows the continuum-subtracted FR551N image (method 1), i.e., the Ly$\alpha$ emission of J1316+2614. As seen in this figure, Ly$\alpha$ is predominantly emitted in the outskirts of the UV-bright stellar core, whose total size is represented by an orange circle with a radius of $2\times r_{\rm eff}$ ($\simeq 0.06^{\prime \prime}$). This bright UV continuum in the center of the galaxy leads to a Ly$\alpha$ hole with weak/residual Ly$\alpha$ emission being emitted at the position of the stellar core. 
It is important to note that the continuum-subtracted Ly$\alpha$ image, especially its faint and diffuse emission, is affected by additional noise due to the subtraction process of the F775W image from the FR551N image.

For consistency, we explore an alternative method for subtracting the stellar contribution in FR551N (method 2 in Figure \ref{fig3}). Using \texttt{PySersic}, we model a Sersic profile with the best-fit parameters obtained for the stellar continuum in F775W (i.e., $r_{\rm eff} = 220$\,pc, $n=2.90$, Table \ref{tab1}) and convolved it with the PSF of FR551N obtained from stars in the field-of-view. After rescaling the flux, we subtract this model from the FR551N image. Consistent with our previous method, we recover the weak/residual Ly$\alpha$ emission within the stellar core. However, we note that the spatial distribution of the bright Ly$\alpha$ emission around the stellar component differs slightly from the previous method, as seen in the bottom panels of Figure \ref{fig3}. Lastly, we investigate the uncertainties on the assumed flux and contribution of the stellar continuum in FR551N. We repeat our analysis, conservatively assuming a stellar contribution in FR551N of $\simeq 40 \%$. Under this assumption, the Ly$\alpha$ "hole" appears less prominent but is still present, with the bulk of Ly$\alpha$ photons emitted around (and far away) from the stellar core.

While the detailed characterization of this hole (e.g., its size) is difficult and requires deeper data, given the low significance of the Ly$\alpha$ emission and other uncertainties in our methodology, our results strongly support a deficit of Ly$\alpha$ co-spatial with the stellar continuum. Such a configuration was discussed and predicted in \cite{marques2022} in order to reconcile the high $f_{\rm esc}^{\rm LyC} \simeq 90\%$, requiring a gas column density $N_{\rm HI} \lesssim 10^{17}$\,cm$^{-2}$, and the large Ly$\alpha$ velocity peak separation observed in J1316+2614 ($\Delta v = 680 \pm 70$\,km\,s$^{-1}$), which suggests  $N_{\rm HI} \sim 10^{21}$\,cm$^{-2}$ according to the radiative models of \cite{verhamme2015} under standard conditions. We further discuss the connection between the Ly$\alpha$ and LyC-UV spatial distributions in Section \ref{resolved_LyC}.

\subsection{Photometry}\label{photometry}

Using \texttt{Sextractor} \citep{bertin2006} we perform aperture photometry on J1316+2614 assuming an aperture of $0.8^{\prime \prime}$ diameter. For F410M, we measure $ \rm F410M = 23.32 \pm 0.06$, which is in excellent agreement with that inferred from the optical GTC/OSIRIS spectrum presented in \cite{marques2022} ($23.33 \pm 0.06$). Similarly, we obtain $ \rm F775W = 21.25 \pm 0.04$ and $ \rm F160W = 21.66 \pm 0.05$ which is consistent with previous ground-based photometry probing similar spectral ranges ($I=21.24\pm0.08$ and $H =21.70 \pm 0.14$, \citealt{marques2022}). For the HAWK-I $K_{\rm s}$-band, we measure $K_{\rm s} = 21.73 \pm 0.06$ which is substantially different than that obtained with GTC/EMIR ($K_{\rm s} = 21.31 \pm 0.07$). These differences are likely due to the shorter spectral coverage of the HAWK-I $K_{\rm s}$-band ($\lambda = 1.984 - 2.308 \mu$m) compared to the GTC/EMIR one ($\lambda = 2.080 - 2.388\mu$m), therefore, not including the contribution of the redshifted [O~{\sc iii}] $\lambda 5008$ emission ($\lambda \simeq 2.310\mu$m) in the former. The ACS/FR551N image of J1316+2614 shows a more extended profile due to the Ly$\alpha$ emission, as shown in Figure \ref{fig2}. Given its extended emission, we use a large aperture of $1.2^{\prime \prime}$ diameter, obtaining $ \rm FR551N = 20.81 \pm 0.10$. Table \ref{tab1} summarizes the photometry of J1316+2614.

In addition to the photometry of the UV-bright and compact starburst, we constrain the flux densities of an underlying stellar population in J1316+2614. So far, the SED analysis by \cite{marques2022} using unresolved photometry supports the absence of a significant old stellar population.  
We assume that the underlying old stellar population has a Gaussian profile with an effective radius of 1.5\,kpc centered at the position of the UV-bright emission. Its size is motivated by the characteristic effective radius of LBGs at similar redshifts \cite[$r_{\rm eff}^{\rm LBGs} \simeq 1.3$\,kpc; e.g.,][]{ribeiro2016} and the size of the dust emission of J1316+2614 detected by ALMA 
($r_{\rm eff}^{\rm dust} \simeq 1.7$\,kpc), which could already be produced before the UV-bright starburst in J1316+2614 (as discussed in Dessauges-Zavadsky et al. in prep.). Given the fact that J1316+2614 is unresolved in the \textit{HST}/F160W and VLT/$K_{\rm s}$ images (section \ref{morphology}), we thus simulate the maximum flux of the underlying stellar population needed to resolve the total emission in these bands. If more extended than the UV-bright component, we find that the underlying old population should be fainter than $\rm F160W > 24.95$ and $K_{\rm s} > 23.98$ to keep the total emission unresolved in F160W and $K_{\rm s}$, respectively.

\subsection{Spectral Energy Distribution}\label{SED}

Using the photometry from the new images along with the previous ones obtained and discussed in \cite{marques2022}, we re-analyze the spectral energy distribution (SED) of J1316+2614. Following \cite{marques2022}, we perform SED-fitting with \texttt{CIGALE} version 2022.1 \citep{Burgarella2005, Boquien2019} using the available photometry and flux measurements of the H$\beta$ and [O~{\sc iii}] $\lambda\lambda$4960,5008. Two stellar components are considered to probe the young UV-bright starburst (assuming a constant star-formation history) and a burst model with an age of 1.4 Gyr to probe the maximum light and mass of an underlying, old stellar population, which corresponds to a formation redshift of $\simeq 12.5$. We assume the \cite{calzetti2000} dust attenuation law and the \cite{chabrier2003} initial mass function. Stellar population models from \cite{bruzual2003} with the metallicity of $Z=0.008$ are considered \citep{marques2022}. We also let $f_{\rm esc}^{\rm LyC}$ as a free parameter. 

\begin{figure}
  \centering
  \includegraphics[width=0.48\textwidth]{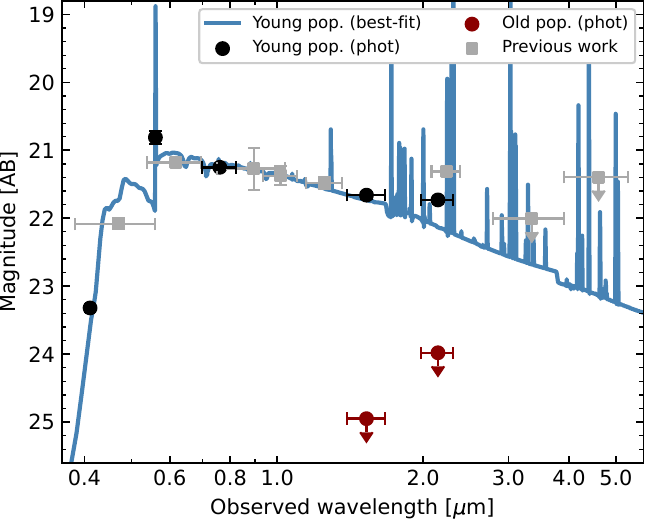}
  \caption{Best-fit SED model (blue) of J1316+2614. The fit uses the new photometry obtained with \textit{HST} and VLT (black circles) and the one presented in \cite{marques2022} (grey squares). The SED of J1316+2614 is dominated by a young stellar population with an age of $5.7 \pm 1.0$\,Myr and a continuous star formation rate $\rm SFR = 898 \pm 181 M_{\odot}$\,yr$^{-1}$. The mass formed in this starburst is log($M_{\star}^{\rm young} / M_{\odot}) = 9.68 \pm 0.03$ with a residual dust attenuation ($E(B-V) = 0.03 \pm 0.01$). The red upper limits represent the maximum flux of the underlying stellar population needed to resolve the emission in the F160W and $K_{\rm s}$ bands (see Section \ref{photometry}).}
  \label{fig_sed}
\end{figure}

Overall, the properties of the UV-bright starburst obtained from \texttt{CIGALE} agree with those previously derived in \cite{marques2022}. Figure \ref{fig_sed} shows the best-fit SED of J1316+2614. The UV-bright starburst is characterized by a young stellar population with an age of $5.7 \pm 1.0$\,Myr and a continuous $\rm SFR = 898 \pm 181 M_{\odot}$\,yr$^{-1}$ with residual dust attenuation ($E(B-V) = 0.03 \pm 0.01$). The SFR derived here reflects the total SFR within 5.7\,Myr, i.e., the age of the young stellar population. If instead we use the 10\,Myr-weighted SFR indicator, we obtain $\rm SFR = 492 \pm 35 \,M_{\odot}$\,yr$^{-1}$ that is consistent with the value reported in \cite{marques2022}, $\rm SFR = 496 \pm 92 \,M_{\odot}$\,yr$^{-1}$. The mass formed in this young starburst is log($M_{\star}^{\rm young} / M_{\odot}) = 9.68 \pm 0.03$, in excellent agreement with previous measurements (log($M_{\star}^{\rm young} / M_{\odot}) = 9.67 \pm 0.07$, \citealt{marques2022}). Our best-fit model also predicts $f_{\rm esc}^{\rm LyC} = 0.75 \pm 0.16$. 

On the other hand, the new and deeper photometry in \textit{HST}/F160W and VLT/$K_{\rm s}$ presented in this work, which probes the rest-optical stellar continuum, provides significantly improved constraints on the properties of the old stellar population.
Our best-fit model predicts an old stellar component, assumed here as a 1.4\,Gyr old burst model, with a stellar mass of log($M_{\star}^{\rm old} / M_{\odot}) = 9.00^{+0.29}_{-1.36}$ or log($M_{\star}^{\rm old} / M_{\odot}) \leq 9.46$ (3$\sigma$). The strong constraints on the mass of the old stellar component are due to the fact that the \textit{HST}/F160W and VLT/$K_{\rm s}$ photometry is fully dominated by the starlight from the young starburst, leaving minimal room for an additional, older stellar component.

In short, our results strongly support that the extremely UV-bright starburst not only dominates the rest-UV and optical light emission of J1316+2614 ($\simeq 100\%$) but also its stellar mass, with a mass fraction of the galaxy formed in the last $\simeq 6$ Myr of $f_{\rm burst} = M_{\star}^{\rm young} / (M_{\star}^{\rm young} + M_{\star}^{\rm old})  \geq 62 \%$ (3$\sigma$). We further discuss the implications of these findings in Section \ref{monolithic}.

\section{Discussion}\label{discussion}

\subsection{J1316+2614 with cluster-like surface densities}\label{disc_cluster}

\begin{figure}
  \centering
  \includegraphics[width=0.41\textwidth]{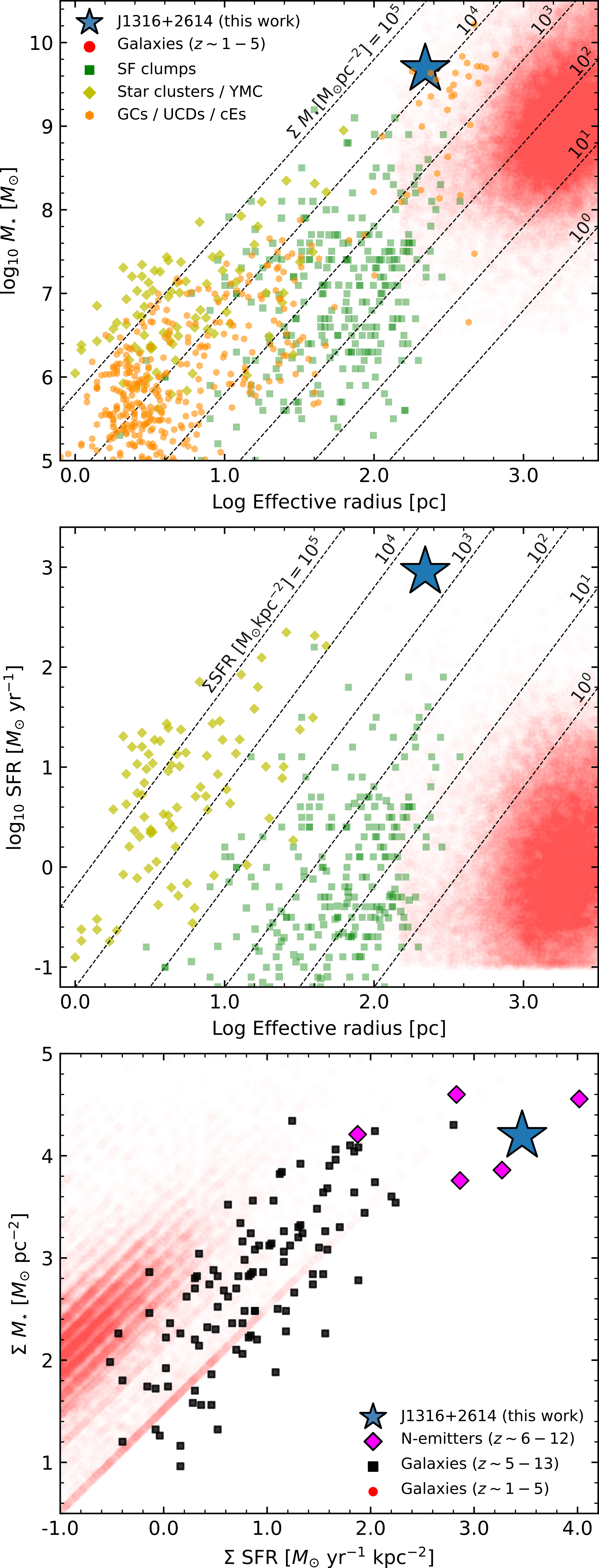}
  \caption{Stellar mass (top) and star-formation rate (middle) as a function of effective radius. J1316+2614 is represented with a blue star. Measurements of other compilations of galaxies at $z=1-5$ \citep[red circles:][]{van2012}, star-forming clumps in lensed galaxies \citep[green squares:][]{Claeyssens2023, Fujimoto2024, Messa2024}, and star-clusters and young massive clusters at different redshifts \citep[yellow diamonds:][]{norris2014, vanzella2023, Adamo2024} are also shown. For star-clusters and star-forming clumps without SFR measurements, we assume star-formation ages of 1\,Myr and 10\,Myr, respectively, and $\rm SFR = M_{\star} / age$. The bottom panel shows the $\Sigma M_{\odot}$ vs. $\Sigma SFR$ of J1316+2614 along with other galaxies at higher redshift (black), including very compact sources with strong nitrogen emission (violet) which exhibit abundance patterns resembling those seen in globular clusters. }
  \label{fig4}
\end{figure}

J1316+2614 is the UV-brightest star-forming galaxy known ($M_{\rm UV} = -24.7$) and one of the most compact. Using the derived mass and SFR from Section \ref{SED}, along with its size ($r_{\rm eff} = 220 \pm 12$\,pc), we measure the stellar mass and SFR surface densities, defined as $\Sigma M_{\star} = M_{\star} / (2 \pi r_{\rm eff}^{2})$ and $\Sigma SFR = SFR / (2 \pi r_{\rm eff}^{2})$. J1316+2614 shows remarkably high mass and SFR surface densities of log($\Sigma M_{\star} [M_{\odot} \rm pc^{-2}]) = 4.20 \pm 0.06$ and log($\Sigma SFR [M_{\odot} \rm yr^{-1} kpc^{-2}]) = 3.47 \pm 0.11$, respectively. Figure \ref{fig4} shows the position of J1316+2614 (blue star) in the mass (top) and SFR (middle) versus $r_{\rm eff}$ diagrams. For comparison, similar measurements are provided for various compilations of galaxies at $z\simeq 1-5$ \citep[red:][]{van2012}, star-forming clumps in lensed galaxies \citep[green:][]{Claeyssens2023, Fujimoto2024, Messa2024}, star-clusters and young massive clusters (YMC) at different redshifts \citep[yellow:][]{norris2014, vanzella2023, Adamo2024}, and local globular clusters (GCs), ultracompact dwarfs (UCDs) and compact elliptical galaxies (cEs) from \cite{norris2014} (orange). 

As shown in Figure \ref{fig4}, the $\Sigma M_{\star}$ and $\Sigma SFR$ of J1316+2614 deviate considerably from those of $z\simeq 1-5$ galaxies, by approximately $1-4$\,dex on average. At these redshifts, star-forming galaxies with high $\Sigma SFR$ are indeed extremely rare, with only very few dusty sub-mm selected galaxies presenting $\Sigma SFR$ values approaching those of J1316+2614 \citep{Oteo2017}. In addition, very few galaxies exhibit similar $\Sigma M_{\star}$, such as UCDs (e.g., M32) and cEs in the local Universe \citep{norris2014}, or extremely massive and compact quiescent galaxies at $z\sim 2-5$ \citep[e.g.,][]{vandokkum2008, barro2017, deGraaff2024, Glazebrook2024}. However, these evolved galaxies have residual star formation. 
At higher redshifts ($z>5$), star-forming galaxies tend to show higher $\Sigma M_{\star}$ and $\Sigma SFR$ than their lower-$z$ counterparts, as recently shown by \cite{Langeroodi2023} and \cite{Morishita2024}. Still, even the densest sources at $z>5$ struggle to reach the densities observed in J1316+2614 (bottom panel in Figure \ref{fig4}). To our knowledge, only a few star-forming galaxies at $z>6$ show comparable densities to J1316+2614 \citep[][Álvarez-Márquez et al. in prep.]{Bunker2023, Williams2023, Castellano2024, Schaerer2024, Topping2024}, several of them are UV-bright and exhibiting peculiar abundance patterns resembling those seen in globular clusters \citep{Charbonnel2023, Marques2024, Schaerer2024, Senchyna2024}. 

The densities derived for J1316+2614 are indeed extreme in star-forming galaxies, and more closely resemble those observed in young massive stellar clusters, which are among the densest systems known (Figure \ref{fig4}). Following \cite{Kruijssen2012}, J1316+2614 would have a very high cluster formation efficiency, $\Gamma \simeq 85\%$, given its high $\Sigma SFR$. 
However, while the surface densities of J1316+2614 are similar to those of massive star clusters, its starburst mass and UV-luminosity differ significantly ($\gtrsim 3-5$\,dex). It remains unclear whether J1316+2614 consists of a large number of normal star clusters ($N\sim 5\times10^{4}$ clusters with $10^{5} M_{\odot}$ each) compacted in a $\sim 220$\,pc radius, or if its luminosity and mass originate from a single, supermassive star cluster with a total mass $M_{\star} \sim 5\times 10^{9} M_{\odot}$.

\subsection{Spatially resolved LyC and gas distribution}\label{resolved_LyC}

\begin{figure*}
  \centering
  \includegraphics[width=0.98\textwidth]{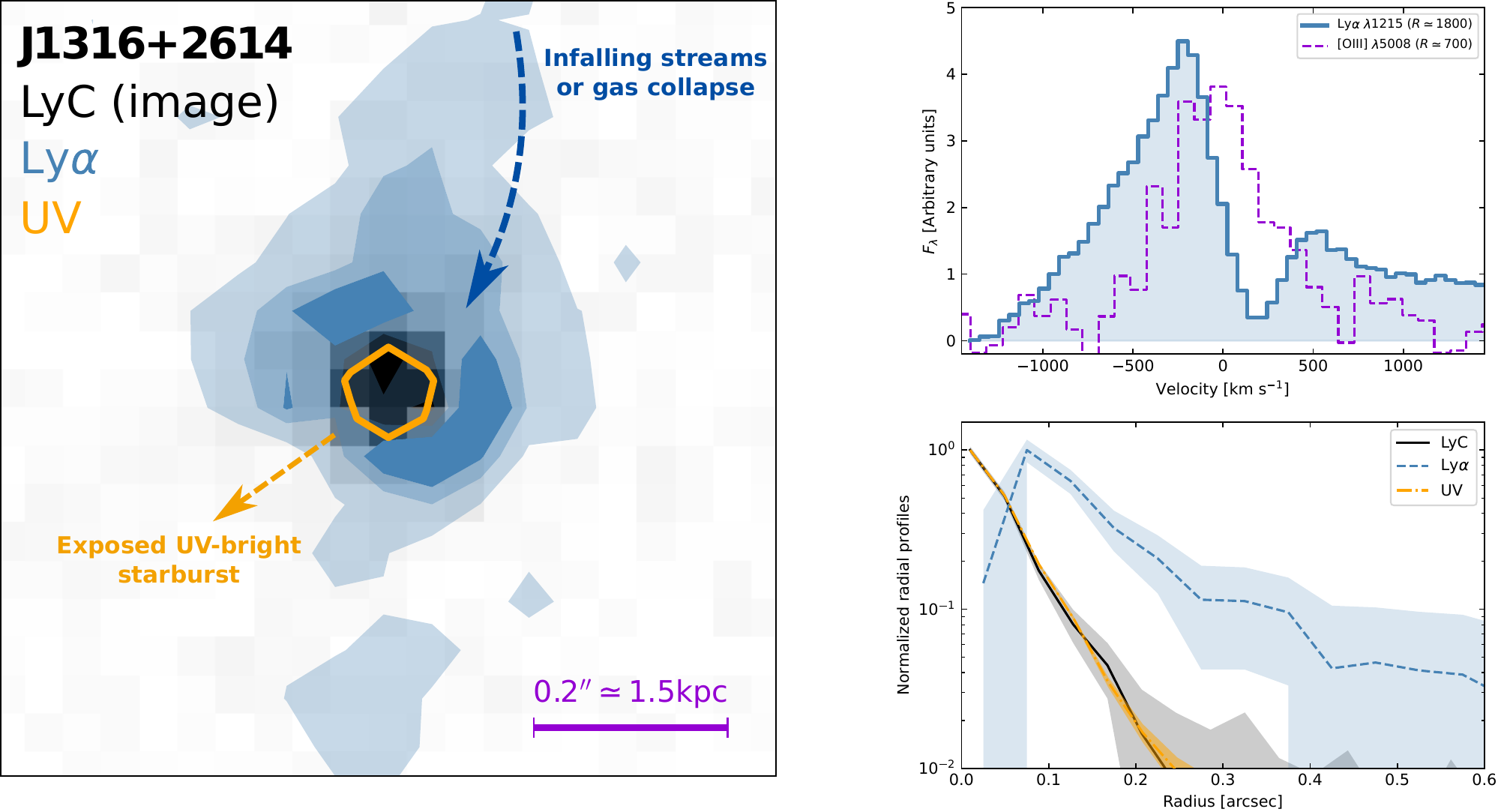}
  \caption{The left panel shows the LyC emission of J1316+2614 from the \textit{HST}/F410M image ($0.8^{\prime \prime} \times 0.8^{\prime \prime}$ size, north is up and east is to the left). The continuum-subtracted Ly$\alpha$ emission is shown in blue with three contours representing the $2.5\sigma-5\sigma$, $5\sigma-8\sigma$, and $8\sigma-15\sigma$ levels. The emission from the non-ionizing UV (F775W, $\lambda_{0} \simeq 1650$\,\AA) is represented in orange and has the size corresponding to its observed FWHM ($\simeq 0.09^{\prime \prime}$). The top right panel shows the GTC optical and near-IR spectra analyzed in \cite{marques2022}, highlighting the Ly$\alpha$ $\lambda$1216\,\AA{ }(blue) and [O~{\sc iii}] $\lambda$5008\,\AA{ }(violet) spectral profile at rest-velocity. The right bottom panel shows the normalized (to their maxima) radial profiles obtained for the LyC (black), UV (orange), and Ly$\alpha$ emission (blue). }
  \label{fig5}
\end{figure*}

J1316+2614 represents the first example of resolved LyC in a strong LyC emitter star-forming galaxy (see also \citealt{Mestric2023} for resolved LyC in a star cluster). As shown in Section \ref{lyc_morphology} and highlighted in Figure \ref{fig5}, the LyC emission is not only resolved but its size and morphology are remarkably similar to that of the non-ionizing UV (with $r_{\rm eff} \simeq 220$\,pc). This is further highlighted in the bottom right panel of Figure \ref{fig5}, where the LyC and UV (normalized) radial profiles are shown (black and yellow, respectively). Despite potential variations in the PSF between F410M and F775W, which are residual (Table \ref{tab1}) and were accounted for in our morphological analysis with \texttt{PySersic} (Section \ref{lyc_morphology}), the figure shows that the LyC and UV radial profiles are indistinguishable within the uncertainties. Together with the high $f_{\rm esc}^{\rm LyC} \approx 90$\% measured in \cite{marques2022}, the almost identical LyC and UV morphologies suggest that the covering fraction of neutral gas and dust, the two known sources of LyC opacity, is residual or negligible. It also suggests that the UV starlight is dominated by O-type stars that emit both LyC and UV photons, which is consistent with the strong wind line profiles seen in the rest-UV spectrum (e.g., N~{\sc iv} $\lambda$1240, C~{\sc iv} $\lambda$1550; \citealt{marques2022}). 
Our results thus support the very high $f_{\rm esc}^{\rm LyC} \simeq 90\%$ directly measured from the optical spectroscopy analyzed in \cite{marques2022}. The non-detection of low-ionization ISM absorption lines in the optical spectrum of J1316+2614 and its steep UV slope ($\beta_{\rm UV} \simeq -2.60$, \citealt{marques2022}) are also consistent with weak/residual gas and dust along the line-of-sight \citep[e.g.,][]{gazagnes2018, chisholm2022, saldana2022}.

The left panel of Figure \ref{fig5} shows the continuum-subtracted Ly$\alpha$ emission obtained from \textit{HST}/FR551N (blue). Ly$\alpha$ photons are predominantly emitted around (and far from) the compact stellar emission as traced by LyC (background image) and UV (orange). This is also highlighted in its radial profile seen in the bottom right panel of Figure \ref{fig5} where Ly$\alpha$ (blue) seems weak within the stellar component ($r < 0.1^{\prime \prime}$). It is worth noting that one of the most puzzling aspects of J1316+2614, as discussed in \cite{marques2022}, is the reconciliation of its high LyC escape fraction with the large Ly$\alpha$ peak separation observed in its spectrum ($\Delta v \simeq 680$\,km\,s$^{-1}$, top right panel of Figure \ref{fig5}). Since density-bounded H~{\sc ii} regions leak LyC photons by $f_{\rm esc}^{\rm LyC} = e^{- \sigma_{\nu_{0}} N_{\rm HI}}$ where $\nu_{0} = 6.3 \times 10^{-18}$\,cm$^{2}$ is the ionization cross-section, the observed $f_{\rm esc} \simeq 90\%$ in J1316+2614 implies a low column density of neutral gas of a few $10^{16}$\,cm$^{-2}$. This contrasts  with the large $\Delta v \rm (Ly\alpha)$ observed in J1316+2614 for which radiative transfer models predict log($N_{\rm HI}/\rm cm^{-2}) \gtrsim 21.5$ under standard assumptions \citep[][]{verhamme2015}. The high $\Delta v$ observed in J1316+2614 can still be reconciled with a low column density of neutral gas of a few $10^{16}$\,cm$^{-2}$, but this requires a fairly high Doppler broadening parameter suggestive of e.g., turbulent gas \citep[][]{Dijkstra2019}. Whether the gas traced by Ly$\alpha$ is optically thick or thin but highly turbulent, this apparent discrepancy seems now solved: Ly$\alpha$ photons are predominantly emitted far from the LyC regions of J1316+2614, and, therefore, the high $f_{\rm esc}^{\rm LyC}$ and large $\Delta v \rm (Ly\alpha)$ can be naturally reconciled. 

Furthermore, the Ly$\alpha$ spectral profile analyzed in \cite{marques2022} also reveals relatively weak emission at the systemic velocity, which should trace the closest gas around the stars. Given that a low column density of neutral gas (of a few $\sim 10^{16}$\,cm$^{-2}$) is necessary for the high $f_{\rm esc}^{\rm LyC} \simeq 90\%$, the weak Ly$\alpha$ emission at the systemic velocity may suggest low amount of ionized gas within and in front the UV stellar component, consistent with the Ly$\alpha$ geometry and its hole shown in Figure \ref{fig5}. A substantial offset between H~{\sc ii} regions and stars in J1316+2614 could also explain its relatively low $O32 = $[O~{\sc iii}] $\lambda$5008 / [O~{\sc ii}] $\lambda$3727 $=4.8 \pm 2.1$ \citep{marques2022} compared to other strong LyC emitters, (e.g., \citealt{jaskot2013, izotov2018b}), and therefore its low ionization parameter ($\propto d^{-2}$, where $d$ is the distance between the ionized gas and stars). This should be confirmed with high-spatial resolution observations of the nebular emission traced by non-resonant lines. 

All the aforementioned points refer to the LyC, UV, and gas distributions along the line-of-sight, and variations of $f_{\rm esc}^{\rm LyC}$ and Ly$\alpha$ properties with sight-line are certainly expected \citep[e.g.,][]{Verhamme2012, mauerhofer2021, Blaizot2023, Gazagnes2024}. However, the analysis by \cite{marques2022} relating the observed H$\beta$ luminosity with the production rate of LyC photons suggests that the total ($\approx 4 \pi$) LyC escape fraction in J1316+2614 is globally high ($f_{\rm esc}^{\rm LyC, 4\pi} \approx 80\%$). 
This conclusion arises from the simple conservation of ionizing photons\footnote{The H$\beta$ luminosity should be proportional to the production rate of ionizing photons, $Q_{\rm H}$, in the form $L(\rm H\beta) \propto$ $Q_{\rm H} \times [1-f_{\rm esc}^{\rm LyC}]$.} and the fact that the non-resonant H$\beta$ emission is weakly affected by sight-line variations, in particular when dust-attenuation levels are residual (as in the case of J1316+2614). 

Finally, we discuss the ALMA observations of J1316+2614 presented in Dessauges-Zavadsky et al. (in prep). J1316+2614 was observed in Bands 3 and 6 to probe the molecular gas and dust emission, respectively. The molecular gas, traced by CO(4-3), is not detected with a $4\sigma$ velocity-integrated intensity limit of $I_{\rm CO} \leq 108$\,mJy\,km\,s$^{-1}$. This places an upper limit on the molecular gas of $M_{\rm molgas} \leq 6.3 \times 10^{9} M_{\odot}$. On the other hand, dust emission is significantly detected (6.2$\sigma$), and a dust mass of $M_{\rm dust} = (3.5 \pm 1.0) \times 10^{7} M_{\odot}$ is inferred (Dessauges-Zavadsky et al. in prep.). As discussed in that work, the origin of the dust remains unclear. It might have formed from SNe in the UV-bright starburst region, although the derived dust mass slightly exceeds standard predictions \citep{gall2018} even without considering dust destruction. The observed dust could also be produced before the UV-bright starburst by an older, still undetected stellar population. Despite the relatively low spatial resolution (beam size of $1.70^{\prime \prime} \times 1.33^{\prime \prime}$), the dust emission is resolved with an effective radius of $r_{\rm eff}^{\rm dust} = 1.7 \pm0.8$\,kpc. If dust and gas are coupled, the dust distribution could follow that traced by Ly$\alpha$, which might explain its relatively large size.
We would expect that the dust emission also shows a ``hole'' or is distributed in a shell, similar to Ly$\alpha$. High angular resolution observations will be needed to test this.
In any case, the steep UV slope ($\beta_{\rm UV} \simeq -2.60$) and the fairly blue ($f_{\nu}$) SED of J1316+2614 indicate residual dust attenuation in the starlight. Thus, it is likely that the dust and stellar emission from the UV-bright starburst are not co-spatial. 

In short, the various independent observations analyzed in this section all indicate that the young, UV-bright starburst in J1316+2614 is likely exposed, i.e., nearly devoid of gas and dust. While gas and dust are present around the starburst, these appear to be residual within it. Under these conditions, LyC photons are free to escape.

\subsection{Energetics and the need for high star-formation efficiency}\label{energy}

The exposed nature of J1316+2614 raises an important question: how can such a vigorous starburst be almost devoid of gas? Although these conditions are extreme on a galaxy-wide scale, they are common in local star clusters and have been studied extensively over the past decade \citep[e.g.,][]{Baumgardt2008, Bastian2014, Krause2016}. The basic principle is that feedback from mechanical and radiative outflows must surpass the gravitational binding force, ejecting the remaining gas from star-forming clouds and leaving an exposed stellar component. In the following, we investigate the different energetic processes associated with J1316+2614.

\begin{figure*}
  \centering
  \includegraphics[width=0.98\textwidth]{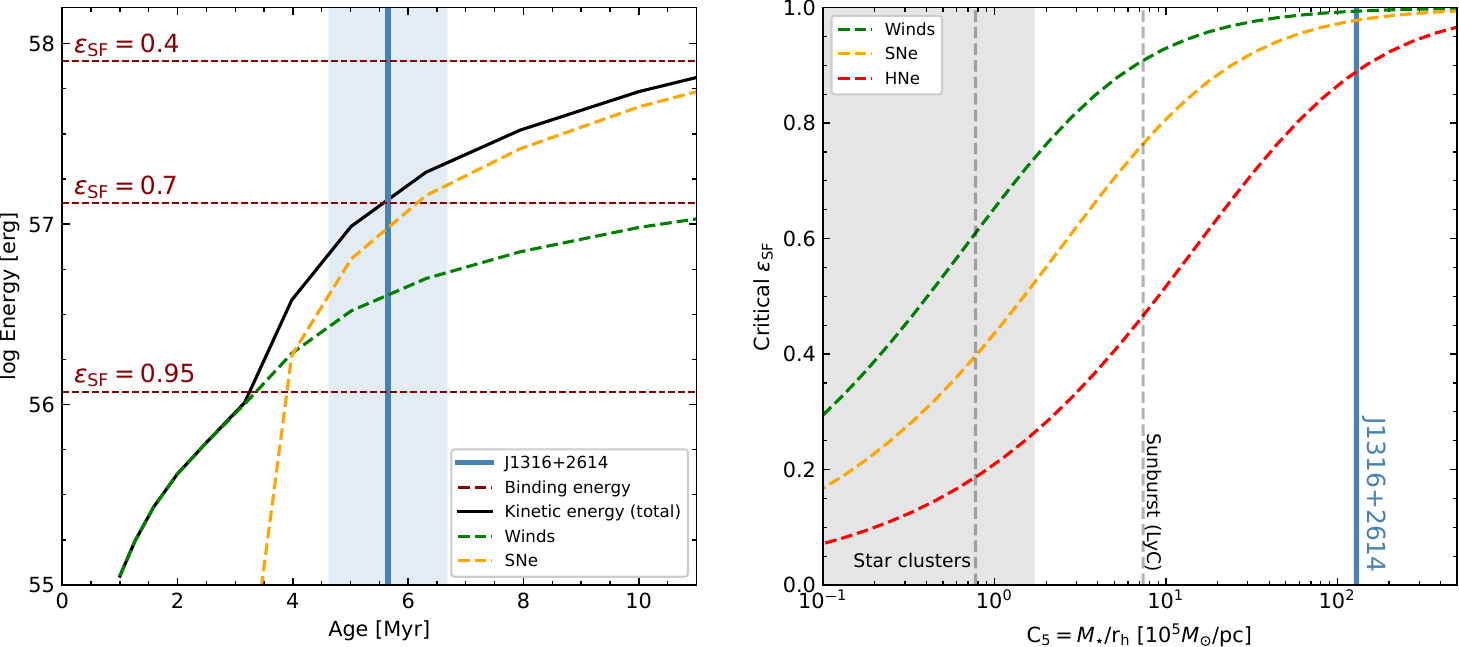}
  \caption{The left panel shows the different energies associated with J1316+2614. Binding energies are shown in dark red (horizontal dashed lines) for star-formation efficiencies of 0.4, 0.7, and 0.95. The total kinetic energy is shown in black and includes the contribution of stellar winds (green dashed line) and supernovae (yellow dashed line). These are obtained from BPASS models assuming a continuous star formation history with $\rm SFR = 898$\,$M_{\odot}$\,yr$^{-1}$, $Z=0.008$ and the \cite{chabrier2003} IMF. The age of J1316+2614 and the corresponding uncertainty are marked in blue. The right panel shows the critical star formation efficiency for gas expulsion by stellar winds (green), normal supernovae ($10^{51}$\,erg, yellow), and hypernovae ($10^{53}$\,erg, red) as a function of the compactness index $C_{5}$ as proposed by \cite{Krause2016}. The location of J1316+2614 is shown in blue. The compactness indexes of other star clusters, including the Sunburst cluster, are marked with grey dashed lines.}
  \label{fig6}
\end{figure*}

We assume the Plummer star cluster model, in which the gravitational binding energy of the gas is given by $E_{\rm b} = k(1-\epsilon_{\rm SF}) G M^{2}_{\rm T} / r_{h}$ \citep{Baumgardt2008}. Here, $k$ is a dimensionless constant ($\approx 0.4$), $\epsilon_{\rm SF}$ is the star formation efficiency, and $M_{\rm T}$ is the total mass enclosed within the half-mass radius $r_{\rm h}$ (where $r_{\rm h} \simeq 1.7\times r_{\rm eff}$ in a Plummer sphere). The factor $(1- \epsilon_{\rm SF})$ accounts for the fact that a fraction of the gas is converted into stars. 
We estimate $E_{\rm b}$ as a function of $\epsilon_{\rm SF}$ assuming that $M_{\rm T} \simeq M_{\star} + M_{\rm gas}$ 
and $\epsilon_{\rm SF} = M_{\star} / M_{\rm T}$. The left panel of Figure \ref{fig6} shows the binding energies for J1316+2614 obtained with $\epsilon_{\rm SF}$ of 0.4, 0.7, and 0.95 (horizontal red lines). For simplicity, we assume that $E_{\rm b}$ remains constant over time. 

We also examine the mechanical energy from stellar winds and supernovae as a function of age. Predictions for the kinematic energy of J1316+2614 were obtained from BPASS v2.2.1 models \citep{stanway2018}, assuming a constant star formation of $\rm SFR=898$\,$M_{\odot}$\,yr$^{-1}$, a metallicity of $Z=0.008$ and the \cite{chabrier2003} IMF, reflecting the properties of the young starburst derived in Section \ref{SED}. The outputs of wind and SN energy of J1316+2614 are shown in the left panel of Figure \ref{fig6} in green and yellow, respectively. As shown, the total mechanical energy (winds and SNe) surpasses the binding energy feedback only if the star-formation efficiency in J1316+2614 is $\epsilon_{\rm SF} \geq 0.7$ (at the age of J1316+2614). In other words, gas expulsion from stellar winds and SNe can only occur if $\epsilon_{\rm SF} \geq 0.7$. 
Before the onset of SNe feedback ($\lesssim 3.5$\,Myr), stellar winds appear relatively inefficient in removing the gas, requiring star-formation efficiencies as high as $\epsilon_{\rm SF} \gtrsim 0.95$.

Given that binding energy is proportional to $M^{2}/r$ and the mechanical energy proportional to $M$, \cite{Krause2016} introduced the compactness index $C_{5} = M_{\odot} / r_{h}$. This index relates the critical star formation efficiency, above which the kinetic energy surpasses the gravitational binding, leading to gas expulsion. The right panel of Figure \ref{fig6} shows the $C_{5}$ derived for J1316+2614, $C_{5} = 128 \pm 11$ ($10^{5} M_{\odot}$\,pc$^{-1}$), along with the critical star formation efficiency needed for gas expulsion by stellar winds (green), supernovae (yellow, $E_{0} = 10^{51}$\,erg) and hypernovae (red, $E_{0} = 10^{53}$\,erg) derived in \cite{Krause2016}. Overall, the $C_{5}$ derived for J1316+2614 is much higher than typical values in star clusters ($C_{5} \lesssim 1$, \citealt{Krause2016}), implying also a much higher $\epsilon_{\rm SF}$ to remove the gas within the stellar core of J1316+2614. However, note that the critical star formation efficiencies shown in this figure were calibrated for star clusters (i.e., assuming a single age burst, \citealt{Krause2016}). Whether J1316+2614 can be described as a single burst is still unclear (see Section \ref{disc_cluster}).

We also explore the effects of radiative-driven outflows, which were recently proposed to explain the overabundance of UV-bright galaxies at high redshifts \citep{Ferrara2023}. 
Following \cite{Ziparo2023}, we determine the conditions under which radiation pressure can drive an outflow by comparing the Eddington ratio as a function of $\Sigma \rm SFR$. Considering $\Sigma \rm SFR \simeq 3\times 10^{3} M_{\odot}$\,yr$^{-1}$\,kpc$^{-2}$ derived for J1316+2614, a radiative-driven outflow can occur when the burstiness parameter is $k_{\rm s} \gtrsim 56$, which quantifies the deviation from the Kennicutt-Schmidt relation (i.e., $\Sigma$SFR $\propto k_{\rm s} \Sigma_{\rm gas}^{1.4}$). This translates to an upper limit of the gas surface density of $\Sigma_{\rm gas} \lesssim 6 \times 10^{3} M_{\odot}$\,pc$^{-2}$. Assuming gas and stars had the same size, we find that $\epsilon_{\rm SF} \gtrsim 0.72$  is needed to launch radiation-driven outflows efficiently. This is consistent with recent radiation hydrodynamic simulations of the formation of massive star clusters, where clusters with $\Sigma M \simeq 10^{3} - 10^{5}$\,$M_{\odot}$\,pc$^{-2}$ became super-Eddington when high star formation efficiencies are reached ($\epsilon_{\rm SF} \sim 80\%$, \citealt{Menon2023}).

Our results thus support a very high star-formation efficiency in J1316+2614. The exposed, gas-free nature of the UV-bright starburst in J1316+2614 suggests that either all the gas was converted into stars ($\epsilon_{\rm SF} \sim 1$) or it was partially ejected by mechanical or radiative feedback, still requiring a fairly high $\epsilon_{\rm SF} \gtrsim 0.7$. Feedback seems indeed ineffective in J1316+2614 to suppress star formation, as seen from the inflowing signatures suggested from its Ly$\alpha$ profile. This inefficient feedback could enhance the star formation efficiency and SFR of J1316+2614, explaining also its remarkably high UV luminosity  \citep[e.g.,][]{Renzini2023}.  
The high star-formation efficiency is also corroborated by the non-detection of molecular gas using ALMA ($M_{\rm gas} \leq 6.3 \times 10^{9} M_{\odot}$), for which an $\epsilon_{\rm SF} \geq 0.4$ was derived (Dessauges-Zavadsky et al. in prep.). 
On the other hand, we acknowledge that these analytic expressions are likely too simple to explain the complex ISM conditions and kinematics of J1316+2614, not considering, for example, possible effects of hot X-ray-emitting gas, turbulence, the multi-phase nature of the ISM, or even the presence of an AGN \citep[see e.g.~][and references therein]{Krause2020, Thompson2024}, although there are currently no signs of such phenomena in our target \citep{marques2022}.
Additionally, the energies and $\epsilon_{\rm SF}$ considered here also depend on the stellar mass derived for J1316+2614, which is sensitive to the adopted IMF \citep[e.g.,][]{Menon2024}. In this context, VMS have been suggested in J1316+2614 (and in other similar UV-bright galaxies, \citealt{Upadhyaya2024}) from its intense and broad He~{\sc ii}~$\lambda 1640$ emission \citep[e.g.,][]{Martins2022, Martins2023}. However, they appear to provide only modest changes on the UV mass-to-light ratio, decreasing it by $\approx \times 1.5$ \citep{Schaerer2024b}.

\subsection{High star-formation efficiency and high LyC escape: cause-effect}

Several surveys have been conducted to understand the conditions under which LyC photons can escape from star-forming galaxies \citep[e.g.,][]{steidel2018, flury2022b}. The standard paradigm assumes that LyC leakage occurs through ionized channels in the ISM originated by strong feedback mechanisms (e.g., \citealt{heckman2001}). 
Observations do suggest the importance of mechanical and radiative driven winds in the escape of LyC photons \citep[e.g.,][]{Komarova2021, Bait2023, Amorin2024, Carr2024}.

In the case of J1316+2614, mechanical and radiative feedback alone seems insufficient to clear the gas within the starburst, at least from a simple energetic balance. 
As highlighted before, gas clearance from mechanical and radiative feedback requires a fairly high star-formation efficiency. Regardless of the presence of strong feedback, an $\epsilon_{\rm SF} \gtrsim 0.7$ is necessary to account for the exposed nature of J1316+2614 and its resulting high LyC leakage. If instead J1316+2614 had a more typical star formation efficiency (e.g., $\lesssim 0.1$), it would likely contain large amounts of gas ($M_{\rm gas} \gtrsim 4 \times 10^{10} M_{\odot}$), increasing considerably the gravitational binding energy ($E_{\rm b} \gtrsim 10^{59}$\,erg). Under these conditions, the kinetic energy ($E_{\rm kin} \simeq 10^{57}$\,erg, Figure \ref{fig6}) would be insufficient to expel the gas from the stellar core, likely preventing LyC leakage. 
Our results thus support that, although mechanical and radiative outflows may be significant in J1316+2614 (though not detected so far), the high star-formation efficiency seems to be the primary driver for the high LyC escape in J1316+2614.

A causal relationship between high star-formation efficiencies and increased LyC production/leakage is indeed expected (c.f., \citealt{Jecmen2023}, \citealt{Kimm2019}, \citealt{Menon2024b}). Simply put, higher $\epsilon_{\rm SF}$ within a star-forming region inevitably results in less residual gas to absorb LyC photons. Additionally, this remaining gas would be more easily removed from the starburst region, as higher $\epsilon_{\rm SF}$ enhances both mechanical and radiative energies (both proportional to SFR or $M_{\star}$, i.e., $\propto \epsilon_{\rm SF} M_{\rm T}$), and decreases the binding energy ($\propto (1- \epsilon_{\rm SF})M_{\rm T}^{2}$). Moreover, an enhanced $\epsilon_{\rm SF}$ would boost the SFR, leading to the formation of more massive stars and thereby increasing LyC emission. This could also lead to density-bounded regions, which facilitates LyC escape \citep[e.g.,][]{Jaskot2017}. Hence, the impact of high $\epsilon_{\rm SF}$ on LyC escape is twofold: it not only enhances the production of ionizing photons but also facilitates their escape. 

While J1316+2614 may represent an extreme case with LyC leakage possibly enhanced by its high star formation efficiency, similar conditions might be already present in other cases. For example, the well-studied, gravitationally lensed Sunburst cluster at $z=2.37$ is known to leak large amounts of LyC photons \citep[][]{dahle2016, rivera2017, rivera2019, vanzella2020}. \cite{Mestric2023} showed that the LyC region is slightly smaller ($r_{\rm eff}^{\rm LyC} \simeq 5$\,pc) than the non-ionizing region ($r_{\rm eff}^{\rm UV} \simeq 8$\,pc), suggesting that the Sunburst cluster is, at least, partially exposed\footnote{It can still be fully exposed if the LyC emitting stars are segregated in the center of the cluster \citep[see the discussion in][]{Mestric2023}.}. Using the stellar mass derived in \cite{Vanzella2022} of $10^{7} M_{\odot}$ and the half-mass radius of $r_{\rm h} \simeq 1.7 \times r_{\rm eff}^{\rm UV}$, the Sunburst cluster shows a high compactness index of $C_{5} \simeq 7.5$ ($10^{5} M_{\odot}$\,pc$^{-1}$), which is a factor of $\approx \times 10$ higher than in local star clusters \citep[e.g.,][]{Bastian2014, Krause2016}. 
If the outflows detected by \cite{Mainali2022} and \cite{Vanzella2022} are responsible for the gas clearance in the Sunburst cluster, then its high $C_{5}$ suggests a high $\epsilon_{\rm SF}$, at least $\gtrsim 0.45$ as seen in the right panel of Figure \ref{fig6}.

In short, our results suggest a close relationship between high $\epsilon_{\rm SF}$ and high escape of ionizing photons. This may be particularly relevant at higher redshifts, where star-formation efficiencies are expected to be higher due to the higher densities and lower metallicities of the ISM in high-$z$ galaxies \citep[e.g.,][]{Dekel2023, Ceverino2024}. Thus, high star-formation efficiencies could not only be crucial for the mass growth of high-$z$ galaxies \citep[e.g.,][]{Xiao2023, deGraaff2024, Glazebrook2024, Weibel2024}, but may also have important implications for cosmic reionization.

\subsection{Feedback-free starburst within an extreme formation mode}\label{monolithic}

\subsubsection{J1316+2614 as an intense feedback-free starburst?}

The high star-formation efficiency in J1316+2614 likely plays a key role in enhancing its star formation rate and burst mass. In this context, high star-formation efficiencies have been recently proposed to explain the high-number density of UV-bright and massive sources at early times \citep{Dekel2023, Li2023, Ceverino2024}. Following \cite{Dekel2023}, high-density environments and low metallicities could favor the formation of so-called feedback-free starbursts (FFB, \citealt{Dekel2023}) through the collapse of gas clouds within very short free-fall times. This would promote higher star formation efficiencies, as the cloud collapse occurs before the onset of mechanical feedback. High-mass galaxies could thus form quickly through several generations of FFBs.

The very high mass density of J1316+2614 indeed suggests a rather short free-fall time. Assuming a spherical gas geometry with a radius of $r_{\rm eff}$, and a star formation efficiency of $0.7$, we derive the free-fall time $t_{\rm ff} \simeq 1.1$\,Myr, where $t_{\rm ff} = \sqrt{3\pi / (32\,G\,\rho)}$ and $\rho = 3\,M_{\rm gas}/(4\pi\,r_{\rm eff}^{3})$. 
Therefore, the derived free-fall time in J1316+2614 is within the range predicted by the FFB scenario ($\sim 1$\,Myr, \citealt{Dekel2023}). Additionally, J1316+2614 shows a very compact morphology ($r_{\rm eff} \simeq 220$\,pc) which aligns with predictions for galaxies in the FFB phase ($r_{\rm eff} \sim 300$\,pc) as discussed in \cite{Li2023}. 

On the other hand, the metallicity inferred for J1316+2614 using the $R23$ method (12+log(O/H)\,$= 8.45 \pm 0.12$, \citealt{marques2022}) is higher than expected in the FFB scenario ($\sim 0.1 Z_{\odot}$, \citealt{Dekel2023}). However, the derived O/H abundance should also reflect the likely efficient chemical enrichment of the starburst itself over the last 5-6 Myr, potentially differing from the gas metallicity in the pre-FFB phase. Furthermore, the derived abundance in J1316+2614 should be treated with caution due to the underlying effect of high $f_{\rm esc}^{\rm LyC}$, as discussed by \cite{marques2022}. 
We also compare the star-formation history of J1316+2614 with those expected in an FFB galaxy \citep{Li2023}. Using the derived $\rm SFR = 898\, M_{\odot}$\,yr$^{-1}$ and the upper limit on the molecular mass $M_{\rm molgas} \leq 6.3\times 10^{9}M_{\odot}$, the gas depletion timescale is $t_{\rm depl} = M_{\rm molgas}/\rm SFR \lesssim 7$\,Myr. Considering its age, the duration of the UV-bright starburst in J1316+2614 should be around $\Delta t \simeq 6-13$\,Myr. The star-formation history of J1316+2614 appears slightly different (higher SFR and higher $\Delta t$) than the predictions by \cite{Li2023} (see their Section 4).

In short, being among the most powerful starbursts known and showing no evidence of feedback so far, J1316+2614 may represent a case of an intense feedback-free starburst with high star-formation efficiency. This would support the link between high star-formation efficiencies and high UV-luminosities, as suggested for galaxies at very high redshifts \citep[][]{Dekel2023, Li2023}. 

\begin{figure*}
  \centering
  \includegraphics[width=0.60\textwidth]{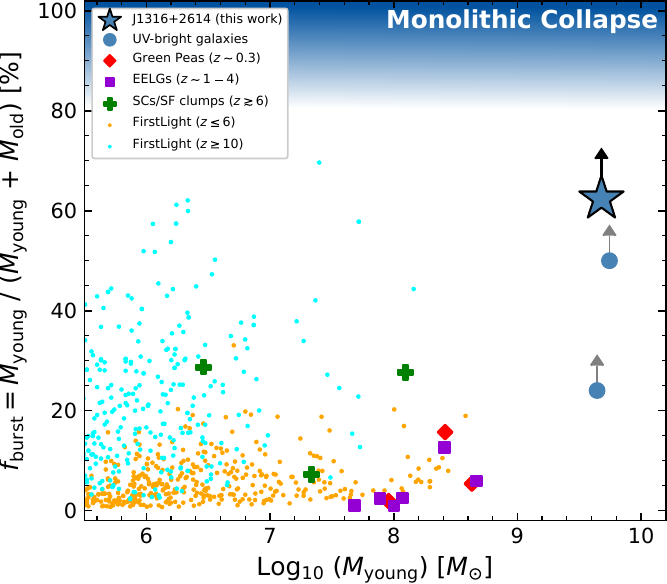}
  \caption{Contribution (in \%) of the starburst mass ($\leq 10$\,Myr) to the total mass of J1316+2614 (blue star) and other starbursting galaxies, including UV-bright galaxies at $z\simeq 2-3$ \citep[blue circles,][]{marques2020b, marques2021}, extreme [O~{\sc iii}] $\lambda$5008 emitters at $z\sim1-4$ \citep[violet squares,][]{Tang2022}, and local Green Pea galaxies \citep[red diamonds,][]{Amorin2012}. Green crosses mark the measurements of highly magnified galaxies at $z\gtrsim 6$ for which young star clusters or star-forming regions are resolved \citep[][]{vanzella2023, Adamo2024, Fujimoto2024}. FirstLight simulated sources at $z\simeq 5-6$ and $z\gtrsim10$ are also shown with orange and cyan dots, respectively \citep{Ceverino2017}.}
  \label{fig8}
\end{figure*}

\subsubsection{Very efficient stellar mass growth}

We further investigate the impact of this extreme UV-bright starburst on the stellar mass growth of J1316+2614. Our multi-wavelength SED analysis in Section \ref{SED} indicates that the young starburst dominates the light (UV-optical) and likely the mass of J1316+2614. Even with conservative assumptions for the old stellar population (1.4 Gyr old), our best-fit \texttt{CIGALE} model predicts a relatively faint stellar component with log($M_{\star}^{\rm old}/M_{\odot}) \leq 9.46$ (3$\sigma$). If so, it could also explain the origin of the dust observed in J1316+2614 ($M_{\rm dust} = 3.5 \times 10^{7} M_{\odot}$, Dessauges-Zavadsky et al. in prep.). Following the $M_{\rm dust} - M_{\star}$ relation derived by \cite{Magnelli2020}, we would expect a stellar mass of the underlying old stellar population of log($M_{\star}^{\rm old}/M_{\odot}) \simeq 9.2$, i.e., consistent with our upper limit. Assuming log($M_{\star}^{\rm old}/M_{\odot}) \leq 9.46$, the UV-bright starburst accounts for a large fraction of the stellar mass of J1316+2614, $f_{\rm burst} = M_{\star}^{\rm young} / (M_{\star}^{\rm young} + M_{\star}^{\rm old}) \geq 62\%$ (3$\sigma$).

Figure \ref{fig8} shows the $f_{\rm burst}$ derived for J1316+2614 (blue star). We also show measurements obtained for other galaxies, including other similar UV-bright galaxies \citep[blue circles,][]{marques2020b, marques2021}, extreme [O~{\sc iii}] $\lambda$5008 emitters at $z\sim1-4$ \citep[squares,][]{Tang2022}, and local Green Pea galaxies \citep[diamonds,][]{Amorin2012}. Green symbols mark the measurements of highly magnified galaxies at $z\gtrsim 6$ for which young star clusters or star-forming regions are resolved \citep[][]{vanzella2023, Adamo2024, Fujimoto2024}.

Green Pea and extreme [O~{\sc iii}] emitters are vigorous star-bursting galaxies, but their young ($\lesssim 10$\,Myr) stellar populations account for a relatively small fraction of the total mass (with a mean a standard deviation of $f_{\rm burst} = 6 \pm 5\%$). This figure also shows the $f_{\rm burst}$ derived for two other UV-bright galaxies, J1220+0842 ($f_{\rm burst} \geq 50\%$, \citealt{marques2020b}) and J0146-0220 ($f_{\rm burst} \geq 24\%$, \citealt{marques2021}), whose properties closely resemble those of J1316+2614 (e.g., $M_{\rm UV} < -24$). Interestingly, Dessauges-Zavadsky et al. (in prep.) find high star formation efficiencies for these two sources, up to $\epsilon_{\rm SF} \geq 24\%$. For more normal, main-sequence galaxies, the $f_{\rm burst}$ parameter is likely even smaller. This is highlighted in Figure \ref{fig8} where we show the predictions from FirstLight simulations \citep{Ceverino2017}. Simulated sources at $z\simeq 5-6$ (orange dots) show relatively modest $f_{\rm burst} \simeq 5\%$ on average. At higher redshifts, FirstLight simulated sources at $z\gtrsim 10$ (cyan dots) show higher stellar mass growth efficiencies, with $f_{\rm burst} \simeq 15\%$ on average, likely due to the more burstier nature of high-$z$ sources and the limited time to form evolved stellar populations. 
Star-forming clumps at $z\sim 6-10$ recently observed in lensed galaxies seem indeed to contribute substantially to the total mass of these early galaxies, up to $f_{\rm burst} \simeq 30\%$ (green in Figure \ref{fig8}, \citealt{vanzella2023, Adamo2024, Fujimoto2024}).

If the star formation efficiency in J1316+2614 is effectively high, our results suggest that it can have a huge impact on the stellar mass growth in the galaxy. Indeed, the high stellar mass fraction formed in the starburst in J1316+2614 ($f_{\rm burst}  \geq 62\%$) resembles traditional models of monolithic collapse, where most of its mass is assembled within a remarkably short period of time \citep[e.g.,][]{Eggen1962, Larson1976, Matteucci1994}.

\subsection{Possible formation paths}

How can J1316+2614 be so UV-bright, compact, massive, and young? Here, we explore possible formation paths for this extreme starburst. Extended Ly$\alpha$ halos, like the one seen in J1316+2614, are commonly observed around high redshift star-forming galaxies \citep[e.g.,][]{leclercq2017, Kusakabe2022}. However, signatures of inflowing gas indicated by blue-dominated Ly$\alpha$ profiles are extremely rare \citep[e.g.,][]{erb2014, martin2015}. Typically, star-forming galaxies show Ly$\alpha$ profiles dominated by redshifted emission along with blueshifted ISM absorption lines, both consistent with large-scale outflows \citep[e.g.,][]{shapley2003,steidel2010, Leclercq2020,marques2020}. Therefore, it is tempting to associate the extreme nature of J1316+2614 with the inflowing gas suggested by its Ly$\alpha$ profile (see discussion in \citealt{marques2022}). 

The rapid mass assembly history of J1316+2614, forming $\simeq 5 \times 10^{9} M_{\odot}$ of stars in just $\simeq 5-6$\,Myr, along with its compactness and the absence of a significant old stellar population, suggests an extreme formation path, potentially monolithic. Inflowing streams or massive gas collapse could trigger and feed the young starburst in J1316+2614. This could provide enough gas supply for a globally high star formation efficiency \citep[e.g., as required for the FFB;][]{Dekel2023}. Thus, the filamentary-like gas distribution traced by Ly$\alpha$ emission (see Figure \ref{fig5}) could represent remnants of the infalling streams, supporting its Ly$\alpha$ spectral profile \citep[e.g.,][]{dijkstra2006}. However, the real extent of this inflowing gas and whether it consists of pristine or recycled material from previous star formation episodes (e.g., galactic fountains) remains unknown.

Alternatively, the inflowing gas kinematics could be related to dissipative compaction of the gas disc induced by a wet-merger or by violent disc instabilities \citep[e.g.,][]{zolotov2015}, as discussed in \cite{marques2022}. However, simulations of the compaction phase predict a modest increase of the sSFR with respect to the main sequence, of $\Delta \rm log  (sSFR) \simeq 0.3-0.7$\,dex \citep{zolotov2015, tacchella2016}, while J1316+2614 shows $\Delta \rm log (sSFR) \simeq 1.7$\,dex (assuming $\rm sSFR = 188 \pm 40 $\,Gyr$^{-1}$). Furthermore, the (stellar) disc remains undetected in our deep, high-resolution images.  
Finally, typical major merger processes involve timescales that are likely too long to explain the observed properties of J1316+2614 \citep[e.g.,][]{Lotz2008}. Given the very young age of J1316+2614, we would expect to observe multiple merging clumps or galaxies. However, \textit{HST} images reveal only a very compact stellar morphology with a half-light radius of $\simeq 220$\,pc. The Ly$\alpha$ emission shows a complex, filamentary-like morphology (oriented south-north, Figures \ref{fig3} and \ref{fig6}), which could, in principle, represent tidal tails from merging galaxies. However, such a configuration is unlikely given the short timescales involved in J1316+2614 ($\simeq 6$\,Myr). Moreover, merging galaxies would likely present evolved stellar populations, but these have not been detected so far in J1316+2614. 
Therefore, a typical major merger seems unlikely to be the direct cause of this intense starburst, though a rare merging configuration cannot be ruled out.

\section{Summary and Conclusions}\label{conclusion}

In this work, we have presented high-resolution Hubble Space Telescope (\textit{HST}) and Very Large Telescope (VLT) imaging observations of J1316+2614 at $z=3.613$ discovered by \cite{marques2022}. J1316+2614 is so far the UV-brightest ($M_{\rm UV} = -24.7$) star-forming galaxy known and one of the strongest LyC emitters with an escape fraction of $f_{\rm esc}^{\rm LyC} \approx 90\%$. It also shows a steep UV slope ($\beta_{\rm UV} \simeq -2.60$) and a peculiar, blue-dominated Ly$\alpha$ emission indicating inflowing gas. The new \textit{HST} observations probe the LyC, Ly$\alpha$, and the rest-UV and optical emission of J1316+2614 with WFC3/F410M, ACS/FR551N, WFC3/F775W, and WFC3/F160W, respectively. Seeing enhanced $K_{\rm s}$-band observations were obtained with VLT/HAWK-I ($\rm FWHM \simeq 0.30^{\prime \prime}$). From the analysis of these data, we arrived at the following main results:

\begin{itemize}
    \item J1316+2614 shows a very compact, but resolved morphology in the LyC and rest-UV. Using \texttt{PySersic} we find similar half-light radii for LyC and UV emission of $r_{\rm eff}^{\rm LyC} = 262 \pm 64$\,pc and $r_{\rm eff}^{\rm UV} = 220 \pm 12$\,pc, respectively. J1316+2614 represents the first case known with resolved LyC emission in a star-forming galaxy. The LyC and UV radial profiles and residuals obtained from the PSF-subtracted images are also indistinguishable within the uncertainties. Our results suggest that the LyC and UV morphologies and sizes are essentially the same ($r_{\rm eff} \simeq 220$\,pc), indicating a residual/null covering fraction of neutral gas and very high LyC leakage. On the other hand, J1316+2614 appears unresolved at longer wavelengths in \textit{HST}/F160W and VLT/$K_{\rm s}$ ($r_{\rm eff}^{\rm opt} \leq 440$\,pc). \vspace{1mm}

    \item The \textit{HST} ACS/FR551N ramp-filter image ($\lambda_{\rm eff}\simeq 5604$\,\AA{ }and width of $\simeq 97$\,\AA), which traces the Ly$\alpha$ emission of J1316+2614, shows a well-resolved morphology with a filamentary-like emission with a total scale length of $\simeq 6.0$\,kpc ($3\sigma$) oriented south-north. After subtracting the contribution of the underlying stellar continuum, Ly$\alpha$ appears residual at the position of the stellar (LyC and UV) emission. Our results indicate a Ly$\alpha$ hole with weak/residual Ly$\alpha$ co-spatial with the stellar continuum. This configuration, combined with its steep UV slope, lack of ISM absorption lines, similar LyC and UV morphologies, and the high LyC escape fraction, suggests that gas and dust are residual within the starburst (though present around it). \vspace{1mm}

    \item Using the photometry obtained from the new images, we re-analyzed its spectral energy distribution (SED). We find that J1316+2614 is dominated by an almost un-obscured ($E(B-V) = 0.03 \pm 0.01$) young stellar population with an age of $5.7 \pm 1.0$\,Myr and a continuous $\rm SFR = 898 \pm 181$\,$M_{\odot}$\,yr$^{-1}$. The mass formed in this young starburst is $M_{\star}^{\rm young} = (4.8 \pm 0.3) \times 10^{9}$\,$M_{\odot}$. The SFR and stellar mass surface densities, log($\Sigma SFR [M_{\odot} \rm yr^{-1} kpc^{-2}]) = 3.47 \pm 0.11$ and log($\Sigma M_{\star} [M_{\odot} \rm pc^{-2}]) = 4.20 \pm 0.06$, are among the highest found in star-forming galaxies, resembling those observed in local young massive star clusters. 
    \vspace{1mm}
    
    \item We also investigated the presence of an underlying old stellar population, which is not detected. Assuming a 1.4 Gyr old burst model, we place an upper limit on its mass of $M_{\star}^{\rm old} \leq 2.8 \times 10^{9} M_{\odot}$ (3$\sigma$). Our results suggest that the UV-bright starburst not only dominates the light emission of J1316+2614 ($\simeq 100\%$) but also its stellar mass, with a mass fraction of the galaxy formed in the last $\simeq 6$ Myr of $f_{\rm burst} = M_{\star}^{\rm young} / (M_{\star}^{\rm young} + M_{\star}^{\rm old})  \geq 62 \%$ (3$\sigma$). Our results suggest that the bulk of the stellar mass in J1316+2614 was assembled within a remarkably short period of time, resembling models of monolithic collapse. 
    
\end{itemize}

The emerging picture of J1316+2614 consists of a very powerful, young, and compact starburst leaking a significant fraction of LyC photons due to the lack of gas and dust within its stellar core. Using simple analytic expressions and assumptions, we explore the different energetic processes associated with J1316+2614 and the conditions leading to its exposed nature. Feedback seems ineffective in J1316+2614 under normal conditions, and a very high star formation efficiency ($\epsilon_{\rm SF} \geq 0.7$) is suggested to expel the remaining gas from the starburst region. Thus, our results support that, although mechanical and radiative outflows may be present in J1316+2614 (though not detected so far), the high star-formation efficiency is likely the main driver for the high LyC escape in J1316+2614.

Overall, the high star formation efficiency in J1316+2614 provides a natural explanation for its remarkably high $f_{\rm esc}^{\rm LyC}$, star formation rate and UV luminosity, and at the same time, the lack of molecular gas ($M_{\rm molgas} \leq 6.3 \times 10^{9} M_{\odot}$, Dessauges-Zavadsky et al in prep.). It also explains the very efficient stellar mass growth in J1316+2614, with at least 62\% of its mass formed in the last 6 Myr. In this context, J1316+2614 may represent an intense feedback-free starburst with enhanced star formation efficiency, similar to those proposed for UV-bright galaxies at very high redshifts. If similar conditions are present in their higher-$z$ counterparts, our results suggest that enhanced star formation efficiencies could be crucial not only for the accelerated mass build-up of high-$z$ galaxies but also for promoting the LyC production and escape, with possible implications for cosmic reionization.

\begin{acknowledgements}
The authors thank the referee for useful comments. 
We would like to thank Angela Adamo and Adélaïde Claeyssens for sharing the data presented in Figure 6. 
This research is based on observations made with the NASA/ESA Hubble Space Telescope obtained from the Space Telescope Science Institute, which is operated by the Association of Universities for Research in Astronomy, Inc., under NASA contract NAS 5–26555. These observations are associated with program ID 17286.
Based on observations collected at the European Southern Observatory under ESO programme 111.251K.001. J.A-M. and L.C. acknowledge support by grant PIB2021-127718NB-100 from the Spanish Ministry of Science and Innovation/State Agency of Research MCIN/AEI/10.13039/501100011033 and by “ERDF A way of making Europe”.

\end{acknowledgements}

%
%

\bibliographystyle{aa}
\bibliography{adssample_v2}

\begin{thebibliography}{124}
\expandafter\ifx\csname natexlab\endcsname\relax\def\natexlab#1{#1}\fi

\bibitem[{{Adamo} {et~al.}(2024){Adamo}, {Bradley}, {Vanzella}, {Claeyssens}, {Welch}, {Diego}, {Mahler}, {Oguri}, {Sharon}, {Abdurro'uf}, {Hsiao}, {Messa}, {Zackrisson}, {Brammer}, {Coe}, {Kokorev}, {Ricotti}, {Zitrin}, {Fujimoto}, {Inoue}, {Resseguier}, {Rigby}, {Jim{\'e}nez-Teja}, {Windhorst}, \& {Xu}}]{Adamo2024}
{Adamo}, A., {Bradley}, L.~D., {Vanzella}, E., {et~al.} 2024, arXiv e-prints, arXiv:2401.03224

\bibitem[{{{\'A}lvarez-M{\'a}rquez} {et~al.}(2021){{\'A}lvarez-M{\'a}rquez}, {Marques-Chaves}, {Colina}, \& {P{\'e}rez-Fournon}}]{alvarez2021}
{{\'A}lvarez-M{\'a}rquez}, J., {Marques-Chaves}, R., {Colina}, L., \& {P{\'e}rez-Fournon}, I. 2021, \aap, 647, A133

\bibitem[{{Amor{\'\i}n} {et~al.}(2012){Amor{\'\i}n}, {P{\'e}rez-Montero}, {V{\'\i}lchez}, \& {Papaderos}}]{Amorin2012}
{Amor{\'\i}n}, R., {P{\'e}rez-Montero}, E., {V{\'\i}lchez}, J.~M., \& {Papaderos}, P. 2012, \apj, 749, 185

\bibitem[{{Amor{\'\i}n} {et~al.}(2024){Amor{\'\i}n}, {Rodr{\'\i}guez-Henr{\'\i}quez}, {Fern{\'a}ndez}, {V{\'\i}lchez}, {Marques-Chaves}, {Schaerer}, {Izotov}, {Firpo}, {Guseva}, {Jaskot}, {Komarova}, {Mu{\~n}oz-Vergara}, {Oey}, {Bait}, {Carr}, {Chisholm}, {Ferguson}, {Flury}, {Giavalisco}, {Hayes}, {Henry}, {Ji}, {King}, {Leclercq}, {{\"O}stlin}, {Pentericci}, {Saldana-Lopez}, {Thuan}, {Trebitsch}, {Wang}, {Worseck}, \& {Xu}}]{Amorin2024}
{Amor{\'\i}n}, R.~O., {Rodr{\'\i}guez-Henr{\'\i}quez}, M., {Fern{\'a}ndez}, V., {et~al.} 2024, \aap, 682, L25

\bibitem[{{Arrabal Haro} {et~al.}(2023){Arrabal Haro}, {Dickinson}, {Finkelstein}, {Fujimoto}, {Fern{\'a}ndez}, {Kartaltepe}, {Jung}, {Cole}, {Burgarella}, {Chworowsky}, {Hutchison}, {Morales}, {Papovich}, {Simons}, {Amor{\'\i}n}, {Backhaus}, {Bagley}, {Bisigello}, {Calabr{\`o}}, {Castellano}, {Cleri}, {Dav{\'e}}, {Dekel}, {Ferguson}, {Fontana}, {Gawiser}, {Giavalisco}, {Harish}, {Hathi}, {Hirschmann}, {Holwerda}, {Huertas-Company}, {Koekemoer}, {Larson}, {Lucas}, {Mobasher}, {P{\'e}rez-Gonz{\'a}lez}, {Pirzkal}, {Rose}, {Santini}, {Trump}, {de la Vega}, {Wang}, {Weiner}, {Wilkins}, {Yang}, {Yung}, \& {Zavala}}]{ArrabalHaro2023}
{Arrabal Haro}, P., {Dickinson}, M., {Finkelstein}, S.~L., {et~al.} 2023, \apjl, 951, L22

\bibitem[{{Bait} {et~al.}(2023){Bait}, {Borthakur}, {Schaerer}, {Momjian}, {Sebastian}, {Saldana-Lopez}, {Flury}, {Chisholm}, {Marques-Chaves}, {Jaskot}, {Ferguson}, {Worseck}, {Ji}, {Komarova}, {Trebitsch}, {Hayes}, {Pentericci}, {Ostlin}, {Thuan}, {Amor{\'\i}n}, {Wang}, {Xu}, \& {Sargent}}]{Bait2023}
{Bait}, O., {Borthakur}, S., {Schaerer}, D., {et~al.} 2023, arXiv e-prints, arXiv:2310.18817

\bibitem[{{Barro} {et~al.}(2017){Barro}, {Faber}, {Koo}, {Dekel}, {Fang}, {Trump}, {P{\'e}rez-Gonz{\'a}lez}, {Pacifici}, {Primack}, {Somerville}, {Yan}, {Guo}, {Liu}, {Ceverino}, {Kocevski}, \& {McGrath}}]{barro2017}
{Barro}, G., {Faber}, S.~M., {Koo}, D.~C., {et~al.} 2017, \apj, 840, 47

\bibitem[{{Bastian} \& {Strader}(2014)}]{Bastian2014}
{Bastian}, N. \& {Strader}, J. 2014, \mnras, 443, 3594

\bibitem[{{Baumgardt} {et~al.}(2008){Baumgardt}, {Kroupa}, \& {Parmentier}}]{Baumgardt2008}
{Baumgardt}, H., {Kroupa}, P., \& {Parmentier}, G. 2008, \mnras, 384, 1231

\bibitem[{{Bertin}(2006)}]{bertin2006}
{Bertin}, E. 2006, in Astronomical Society of the Pacific Conference Series, Vol. 351, Astronomical Data Analysis Software and Systems XV, ed. C.~{Gabriel}, C.~{Arviset}, D.~{Ponz}, \& S.~{Enrique}, 112

\bibitem[{{Birrer} {et~al.}(2022){Birrer}, {Bhamre}, {Nierenberg}, {Yang}, \& {Van de Vyvere}}]{Birrer2022}
{Birrer}, S., {Bhamre}, V., {Nierenberg}, A., {Yang}, L., \& {Van de Vyvere}, L. 2022, {PSFr: Point Spread Function reconstruction}, Astrophysics Source Code Library, record ascl:2210.005

\bibitem[{{Birrer} {et~al.}(2021){Birrer}, {Shajib}, {Gilman}, {Galan}, {Aalbers}, {Millon}, {Morgan}, {Pagano}, {Park}, {Teodori}, {Tessore}, {Ueland}, {Van de Vyvere}, {Wagner-Carena}, {Wempe}, {Yang}, {Ding}, {Schmidt}, {Sluse}, {Zhang}, \& {Amara}}]{Birrer2021}
{Birrer}, S., {Shajib}, A., {Gilman}, D., {et~al.} 2021, The Journal of Open Source Software, 6, 3283

\bibitem[{{Blaizot} {et~al.}(2023){Blaizot}, {Garel}, {Verhamme}, {Katz}, {Kimm}, {Michel-Dansac}, {Mitchell}, {Rosdahl}, \& {Trebitsch}}]{Blaizot2023}
{Blaizot}, J., {Garel}, T., {Verhamme}, A., {et~al.} 2023, \mnras, 523, 3749

\bibitem[{{Boquien} {et~al.}(2019){Boquien}, {Burgarella}, {Roehlly}, {Buat}, {Ciesla}, {Corre}, {Inoue}, \& {Salas}}]{Boquien2019}
{Boquien}, M., {Burgarella}, D., {Roehlly}, Y., {et~al.} 2019, \aap, 622, A103

\bibitem[{{Bouwens} {et~al.}(2021){Bouwens}, {Oesch}, {Stefanon}, {Illingworth}, {Labb{\'e}}, {Reddy}, {Atek}, {Montes}, {Naidu}, {Nanayakkara}, {Nelson}, \& {Wilkins}}]{Bouwens2021}
{Bouwens}, R.~J., {Oesch}, P.~A., {Stefanon}, M., {et~al.} 2021, \aj, 162, 47

\bibitem[{{Boylan-Kolchin}(2024)}]{Boylan-Kolchin2024}
{Boylan-Kolchin}, M. 2024, arXiv e-prints, arXiv:2407.10900

\bibitem[{{Bruzual} \& {Charlot}(2003)}]{bruzual2003}
{Bruzual}, G. \& {Charlot}, S. 2003, \mnras, 344, 1000

\bibitem[{{Bunker} {et~al.}(2023){Bunker}, {Saxena}, {Cameron}, {Willott}, {Curtis-Lake}, {Jakobsen}, {Carniani}, {Smit}, {Maiolino}, {Witstok}, {Curti}, {D'Eugenio}, {Jones}, {Ferruit}, {Arribas}, {Charlot}, {Chevallard}, {Giardino}, {de Graaff}, {Looser}, {L{\"u}tzgendorf}, {Maseda}, {Rawle}, {Rix}, {Del Pino}, {Alberts}, {Egami}, {Eisenstein}, {Endsley}, {Hainline}, {Hausen}, {Johnson}, {Rieke}, {Rieke}, {Robertson}, {Shivaei}, {Stark}, {Sun}, {Tacchella}, {Tang}, {Williams}, {Willmer}, {Baker}, {Baum}, {Bhatawdekar}, {Bowler}, {Boyett}, {Chen}, {Circosta}, {Helton}, {Ji}, {Kumari}, {Lyu}, {Nelson}, {Parlanti}, {Perna}, {Sandles}, {Scholtz}, {Suess}, {Topping}, {{\"U}bler}, {Wallace}, \& {Whitler}}]{Bunker2023}
{Bunker}, A.~J., {Saxena}, A., {Cameron}, A.~J., {et~al.} 2023, \aap, 677, A88

\bibitem[{{Burgarella} {et~al.}(2005){Burgarella}, {Buat}, \& {Iglesias-P\'aramo}}]{Burgarella2005}
{Burgarella}, D., {Buat}, V., \& {Iglesias-P\'aramo}, J. 2005, \mnras, 360, 1413

\bibitem[{{Calzetti} {et~al.}(2000){Calzetti}, {Armus}, {Bohlin}, {Kinney}, {Koornneef}, \& {Storchi-Bergmann}}]{calzetti2000}
{Calzetti}, D., {Armus}, L., {Bohlin}, R.~C., {et~al.} 2000, \apj, 533, 682

\bibitem[{{Carniani} {et~al.}(2024){Carniani}, {Hainline}, {D'Eugenio}, {Eisenstein}, {Jakobsen}, {Witstok}, {Johnson}, {Chevallard}, {Maiolino}, {Helton}, {Willott}, {Robertson}, {Alberts}, {Arribas}, {Baker}, {Bhatawdekar}, {Boyett}, {Bunker}, {Cameron}, {Cargile}, {Charlot}, {Curti}, {Curtis-Lake}, {Egami}, {Giardino}, {Isaak}, {Ji}, {Jones}, {Maseda}, {Parlanti}, {Rawle}, {Rieke}, {Rieke}, {Rodr{\'\i}guez Del Pino}, {Saxena}, {Scholtz}, {Smit}, {Sun}, {Tacchella}, {{\"U}bler}, {Venturi}, {Williams}, \& {Willmer}}]{Carniani2024}
{Carniani}, S., {Hainline}, K., {D'Eugenio}, F., {et~al.} 2024, arXiv e-prints, arXiv:2405.18485

\bibitem[{{Carr} {et~al.}(2024){Carr}, {Cen}, {Scarlata}, {Xu}, {Henry}, {Marques-Chaves}, {Schaerer}, {Amor{\'\i}n}, {Oey}, {Komarova}, {Flury}, {Jaskot}, {Saldana-Lopez}, {Ji}, {Huberty}, {Heckman}, {Ostlin}, {Bait}, {Hayes}, {Thuan}, {Berg}, {Giavalisco}, {Borthakur}, {Chisholm}, {Ferguson}, {Michel-Dansac}, {Verhamme}, \& {Worseck}}]{Carr2024}
{Carr}, C.~A., {Cen}, R., {Scarlata}, C., {et~al.} 2024, arXiv e-prints, arXiv:2409.05180

\bibitem[{{Castellano} {et~al.}(2024){Castellano}, {Napolitano}, {Fontana}, {Roberts-Borsani}, {Treu}, {Vanzella}, {Zavala}, {Arrabal Haro}, {Calabr{\`o}}, {Llerena}, {Mascia}, {Merlin}, {Paris}, {Pentericci}, {Santini}, {Bakx}, {Bergamini}, {Cupani}, {Dickinson}, {Filippenko}, {Glazebrook}, {Grillo}, {Kelly}, {Malkan}, {Mason}, {Morishita}, {Nanayakkara}, {Rosati}, {Sani}, {Wang}, \& {Yoon}}]{Castellano2024}
{Castellano}, M., {Napolitano}, L., {Fontana}, A., {et~al.} 2024, arXiv e-prints, arXiv:2403.10238

\bibitem[{{Ceverino} {et~al.}(2017){Ceverino}, {Glover}, \& {Klessen}}]{Ceverino2017}
{Ceverino}, D., {Glover}, S. C.~O., \& {Klessen}, R.~S. 2017, \mnras, 470, 2791

\bibitem[{{Ceverino} {et~al.}(2024){Ceverino}, {Nakazato}, {Yoshida}, {Klessen}, \& {Glover}}]{Ceverino2024}
{Ceverino}, D., {Nakazato}, Y., {Yoshida}, N., {Klessen}, R., \& {Glover}, S. 2024, arXiv e-prints, arXiv:2404.02537

\bibitem[{{Chabrier}(2003)}]{chabrier2003}
{Chabrier}, G. 2003, \apjl, 586, L133

\bibitem[{{Charbonnel} {et~al.}(2023){Charbonnel}, {Schaerer}, {Prantzos}, {Ram{\'\i}rez-Galeano}, {Fragos}, {Kuruvanthodi}, {Marques-Chaves}, \& {Gieles}}]{Charbonnel2023}
{Charbonnel}, C., {Schaerer}, D., {Prantzos}, N., {et~al.} 2023, \aap, 673, L7

\bibitem[{{Chisholm} {et~al.}(2022){Chisholm}, {Saldana-Lopez}, {Flury}, {Schaerer}, {Jaskot}, {Amor{\'\i}n}, {Atek}, {Finkelstein}, {Fleming}, {Ferguson}, {Fern{\'a}ndez}, {Giavalisco}, {Hayes}, {Heckman}, {Henry}, {Ji}, {Marques-Chaves}, {Mauerhofer}, {McCandliss}, {Oey}, {{\"O}stlin}, {Rutkowski}, {Scarlata}, {Thuan}, {Trebitsch}, {Wang}, {Worseck}, \& {Xu}}]{chisholm2022}
{Chisholm}, J., {Saldana-Lopez}, A., {Flury}, S., {et~al.} 2022, \mnras, 517, 5104

\bibitem[{{Claeyssens} {et~al.}(2023){Claeyssens}, {Adamo}, {Richard}, {Mahler}, {Messa}, \& {Dessauges-Zavadsky}}]{Claeyssens2023}
{Claeyssens}, A., {Adamo}, A., {Richard}, J., {et~al.} 2023, \mnras, 520, 2180

\bibitem[{{Dahle} {et~al.}(2016){Dahle}, {Aghanim}, {Guennou}, {Hudelot}, {Kneissl}, {Pointecouteau}, {Beelen}, {Bayliss}, {Douspis}, {Nesvadba}, {Hempel}, {Gronke}, {Burenin}, {Dole}, {Harrison}, {Mazzotta}, \& {Sunyaev}}]{dahle2016}
{Dahle}, H., {Aghanim}, N., {Guennou}, L., {et~al.} 2016, \aap, 590, L4

\bibitem[{{de Graaff} {et~al.}(2024){de Graaff}, {Setton}, {Brammer}, {Cutler}, {Suess}, {Labbe}, {Leja}, {Weibel}, {Maseda}, {Whitaker}, {Bezanson}, {Boogaard}, {Cleri}, {De Lucia}, {Franx}, {Greene}, {Hirschmann}, {Matthee}, {McConachie}, {Naidu}, {Oesch}, {Price}, {Rix}, {Valentino}, {Wang}, \& {Williams}}]{deGraaff2024}
{de Graaff}, A., {Setton}, D.~J., {Brammer}, G., {et~al.} 2024, arXiv e-prints, arXiv:2404.05683

\bibitem[{{Dekel} {et~al.}(2023){Dekel}, {Sarkar}, {Birnboim}, {Mandelker}, \& {Li}}]{Dekel2023}
{Dekel}, A., {Sarkar}, K.~C., {Birnboim}, Y., {Mandelker}, N., \& {Li}, Z. 2023, \mnras, 523, 3201

\bibitem[{{Dijkstra}(2019)}]{Dijkstra2019}
{Dijkstra}, M. 2019, Saas-Fee Advanced Course, 46, 1

\bibitem[{{Dijkstra} {et~al.}(2006){Dijkstra}, {Haiman}, \& {Spaans}}]{dijkstra2006}
{Dijkstra}, M., {Haiman}, Z., \& {Spaans}, M. 2006, \apj, 649, 14

\bibitem[{{Eggen} {et~al.}(1962){Eggen}, {Lynden-Bell}, \& {Sandage}}]{Eggen1962}
{Eggen}, O.~J., {Lynden-Bell}, D., \& {Sandage}, A.~R. 1962, \apj, 136, 748

\bibitem[{{Erb} {et~al.}(2014){Erb}, {Steidel}, {Trainor}, {Bogosavljevi{\'c}}, {Shapley}, {Nestor}, {Kulas}, {Law}, {Strom}, {Rudie}, {Reddy}, {Pettini}, {Konidaris}, {Mace}, {Matthews}, \& {McLean}}]{erb2014}
{Erb}, D.~K., {Steidel}, C.~C., {Trainor}, R.~F., {et~al.} 2014, \apj, 795, 33

\bibitem[{{Ferrara} {et~al.}(2023){Ferrara}, {Pallottini}, \& {Dayal}}]{Ferrara2023}
{Ferrara}, A., {Pallottini}, A., \& {Dayal}, P. 2023, \mnras, 522, 3986

\bibitem[{{Finkelstein} {et~al.}(2023){Finkelstein}, {Bagley}, {Ferguson}, {Wilkins}, {Kartaltepe}, {Papovich}, {Yung}, {Arrabal Haro}, {Behroozi}, {Dickinson}, {Kocevski}, {Koekemoer}, {Larson}, {Le Bail}, {Morales}, {P{\'e}rez-Gonz{\'a}lez}, {Burgarella}, {Dav{\'e}}, {Hirschmann}, {Somerville}, {Wuyts}, {Bromm}, {Casey}, {Fontana}, {Fujimoto}, {Gardner}, {Giavalisco}, {Grazian}, {Grogin}, {Hathi}, {Hutchison}, {Jha}, {Jogee}, {Kewley}, {Kirkpatrick}, {Long}, {Lotz}, {Pentericci}, {Pierel}, {Pirzkal}, {Ravindranath}, {Ryan}, {Trump}, {Yang}, {Bhatawdekar}, {Bisigello}, {Buat}, {Calabr{\`o}}, {Castellano}, {Cleri}, {Cooper}, {Croton}, {Daddi}, {Dekel}, {Elbaz}, {Franco}, {Gawiser}, {Holwerda}, {Huertas-Company}, {Jaskot}, {Leung}, {Lucas}, {Mobasher}, {Pandya}, {Tacchella}, {Weiner}, \& {Zavala}}]{Finkelstein2023}
{Finkelstein}, S.~L., {Bagley}, M.~B., {Ferguson}, H.~C., {et~al.} 2023, \apjl, 946, L13

\bibitem[{{Flury} {et~al.}(2022){Flury}, {Jaskot}, {Ferguson}, {Worseck}, {Makan}, {Chisholm}, {Saldana-Lopez}, {Schaerer}, {McCandliss}, {Xu}, {Wang}, {Oey}, {Ford}, {Heckman}, {Ji}, {Giavalisco}, {Amor{\'\i}n}, {Atek}, {Blaizot}, {Borthakur}, {Carr}, {Castellano}, {Barros}, {Dickinson}, {Finkelstein}, {Fleming}, {Fontanot}, {Garel}, {Grazian}, {Hayes}, {Henry}, {Mauerhofer}, {Micheva}, {Ostlin}, {Papovich}, {Pentericci}, {Ravindranath}, {Rosdahl}, {Rutkowski}, {Santini}, {Scarlata}, {Teplitz}, {Thuan}, {Trebitsch}, {Vanzella}, \& {Verhamme}}]{flury2022b}
{Flury}, S.~R., {Jaskot}, A.~E., {Ferguson}, H.~C., {et~al.} 2022, \apj, 930, 126

\bibitem[{{Fruchter} \& {Hook}(2002)}]{Fruchter2002}
{Fruchter}, A.~S. \& {Hook}, R.~N. 2002, \pasp, 114, 144

\bibitem[{{Fujimoto} {et~al.}(2024){Fujimoto}, {Ouchi}, {Kohno}, {Valentino}, {Gim{\'e}nez-Arteaga}, {Brammer}, {Furtak}, {Kohandel}, {Oguri}, {Pallottini}, {Richard}, {Zitrin}, {Bauer}, {Boylan-Kolchin}, {Dessauges-Zavadsky}, {Egami}, {Finkelstein}, {Ma}, {Smail}, {Watson}, {Hutchison}, {Rigby}, {Welch}, {Ao}, {Bradley}, {Caminha}, {Caputi}, {Espada}, {Endsley}, {Fudamoto}, {Gonz{\'a}lez-L{\'o}pez}, {Hatsukade}, {Koekemoer}, {Kokorev}, {Laporte}, {Lee}, {Magdis}, {Ono}, {Rizzo}, {Shibuya}, {Shimasaku}, {Sun}, {Toft}, {Umehata}, {Wang}, \& {Yajima}}]{Fujimoto2024}
{Fujimoto}, S., {Ouchi}, M., {Kohno}, K., {et~al.} 2024, arXiv e-prints, arXiv:2402.18543

\bibitem[{{Gaia Collaboration} {et~al.}(2023){Gaia Collaboration}, {Vallenari}, {Brown}, {Prusti}, {de Bruijne}, {Arenou}, {Babusiaux}, {Biermann}, {Creevey}, {Ducourant}, {Evans}, {Eyer}, {Guerra}, {Hutton}, {Jordi}, {Klioner}, {Lammers}, {Lindegren}, {Luri}, {Mignard}, {Panem}, {Pourbaix}, {Randich}, {Sartoretti}, {Soubiran}, {Tanga}, {Walton}, {Bailer-Jones}, {Bastian}, {Drimmel}, {Jansen}, {Katz}, {Lattanzi}, {van Leeuwen}, {Bakker}, {Cacciari}, {Casta{\~n}eda}, {De Angeli}, {Fabricius}, {Fouesneau}, {Fr{\'e}mat}, {Galluccio}, {Guerrier}, {Heiter}, {Masana}, {Messineo}, {Mowlavi}, {Nicolas}, {Nienartowicz}, {Pailler}, {Panuzzo}, {Riclet}, {Roux}, {Seabroke}, {Sordo}, {Th{\'e}venin}, {Gracia-Abril}, {Portell}, {Teyssier}, {Altmann}, {Andrae}, {Audard}, {Bellas-Velidis}, {Benson}, {Berthier}, {Blomme}, {Burgess}, {Busonero}, {Busso}, {C{\'a}novas}, {Carry}, {Cellino}, {Cheek}, {Clementini}, {Damerdji}, {Davidson}, {de Teodoro}, {Nu{\~n}ez Campos}, {Delchambre}, {Dell'Oro}, {Esquej},
  {Fern{\'a}ndez-Hern{\'a}ndez}, {Fraile}, {Garabato}, {Garc{\'\i}a-Lario}, {Gosset}, {Haigron}, {Halbwachs}, {Hambly}, {Harrison}, {Hern{\'a}ndez}, {Hestroffer}, {Hodgkin}, {Holl}, {Jan{\ss}en}, {Jevardat de Fombelle}, {Jordan}, {Krone-Martins}, {Lanzafame}, {L{\"o}ffler}, {Marchal}, {Marrese}, {Moitinho}, {Muinonen}, {Osborne}, {Pancino}, {Pauwels}, {Recio-Blanco}, {Reyl{\'e}}, {Riello}, {Rimoldini}, {Roegiers}, {Rybizki}, {Sarro}, {Siopis}, {Smith}, {Sozzetti}, {Utrilla}, {van Leeuwen}, {Abbas}, {{\'A}brah{\'a}m}, {Abreu Aramburu}, {Aerts}, {Aguado}, {Ajaj}, {Aldea-Montero}, {Altavilla}, {{\'A}lvarez}, {Alves}, {Anders}, {Anderson}, {Anglada Varela}, {Antoja}, {Baines}, {Baker}, {Balaguer-N{\'u}{\~n}ez}, {Balbinot}, {Balog}, {Barache}, {Barbato}, {Barros}, {Barstow}, {Bartolom{\'e}}, {Bassilana}, {Bauchet}, {Becciani}, {Bellazzini}, {Berihuete}, {Bernet}, {Bertone}, {Bianchi}, {Binnenfeld}, {Blanco-Cuaresma}, {Blazere}, {Boch}, {Bombrun}, {Bossini}, {Bouquillon}, {Bragaglia}, {Bramante}, {Breedt},
  {Bressan}, {Brouillet}, {Brugaletta}, {Bucciarelli}, {Burlacu}, {Butkevich}, {Buzzi}, {Caffau}, {Cancelliere}, {Cantat-Gaudin}, {Carballo}, {Carlucci}, {Carnerero}, {Carrasco}, {Casamiquela}, {Castellani}, {Castro-Ginard}, {Chaoul}, {Charlot}, {Chemin}, {Chiaramida}, {Chiavassa}, {Chornay}, {Comoretto}, {Contursi}, {Cooper}, {Cornez}, {Cowell}, {Crifo}, {Cropper}, {Crosta}, {Crowley}, {Dafonte}, {Dapergolas}, {David}, {David}, {de Laverny}, {De Luise}, {De March}, {De Ridder}, {de Souza}, {de Torres}, {del Peloso}, {del Pozo}, {Delbo}, {Delgado}, {Delisle}, {Demouchy}, {Dharmawardena}, {Di Matteo}, {Diakite}, {Diener}, {Distefano}, {Dolding}, {Edvardsson}, {Enke}, {Fabre}, {Fabrizio}, {Faigler}, {Fedorets}, {Fernique}, {Fienga}, {Figueras}, {Fournier}, {Fouron}, {Fragkoudi}, {Gai}, {Garcia-Gutierrez}, {Garcia-Reinaldos}, {Garc{\'\i}a-Torres}, {Garofalo}, {Gavel}, {Gavras}, {Gerlach}, {Geyer}, {Giacobbe}, {Gilmore}, {Girona}, {Giuffrida}, {Gomel}, {Gomez}, {Gonz{\'a}lez-N{\'u}{\~n}ez},
  {Gonz{\'a}lez-Santamar{\'\i}a}, {Gonz{\'a}lez-Vidal}, {Granvik}, {Guillout}, {Guiraud}, {Guti{\'e}rrez-S{\'a}nchez}, {Guy}, {Hatzidimitriou}, {Hauser}, {Haywood}, {Helmer}, {Helmi}, {Sarmiento}, {Hidalgo}, {Hilger}, {H{\l}adczuk}, {Hobbs}, {Holland}, {Huckle}, {Jardine}, {Jasniewicz}, {Jean-Antoine Piccolo}, {Jim{\'e}nez-Arranz}, {Jorissen}, {Juaristi Campillo}, {Julbe}, {Karbevska}, {Kervella}, {Khanna}, {Kontizas}, {Kordopatis}, {Korn}, {K{\'o}sp{\'a}l}, {Kostrzewa-Rutkowska}, {Kruszy{\'n}ska}, {Kun}, {Laizeau}, {Lambert}, {Lanza}, {Lasne}, {Le Campion}, {Lebreton}, {Lebzelter}, {Leccia}, {Leclerc}, {Lecoeur-Taibi}, {Liao}, {Licata}, {Lindstr{\o}m}, {Lister}, {Livanou}, {Lobel}, {Lorca}, {Loup}, {Madrero Pardo}, {Magdaleno Romeo}, {Managau}, {Mann}, {Manteiga}, {Marchant}, {Marconi}, {Marcos}, {Marcos Santos}, {Mar{\'\i}n Pina}, {Marinoni}, {Marocco}, {Marshall}, {Martin Polo}, {Mart{\'\i}n-Fleitas}, {Marton}, {Mary}, {Masip}, {Massari}, {Mastrobuono-Battisti}, {Mazeh}, {McMillan}, {Messina}, {Michalik},
  {Millar}, {Mints}, {Molina}, {Molinaro}, {Moln{\'a}r}, {Monari}, {Mongui{\'o}}, {Montegriffo}, {Montero}, {Mor}, {Mora}, {Morbidelli}, {Morel}, {Morris}, {Muraveva}, {Murphy}, {Musella}, {Nagy}, {Noval}, {Oca{\~n}a}, {Ogden}, {Ordenovic}, {Osinde}, {Pagani}, {Pagano}, {Palaversa}, {Palicio}, {Pallas-Quintela}, {Panahi}, {Payne-Wardenaar}, {Pe{\~n}alosa Esteller}, {Penttil{\"a}}, {Pichon}, {Piersimoni}, {Pineau}, {Plachy}, {Plum}, {Poggio}, {Pr{\v{s}}a}, {Pulone}, {Racero}, {Ragaini}, {Rainer}, {Raiteri}, {Rambaux}, {Ramos}, {Ramos-Lerate}, {Re Fiorentin}, {Regibo}, {Richards}, {Rios Diaz}, {Ripepi}, {Riva}, {Rix}, {Rixon}, {Robichon}, {Robin}, {Robin}, {Roelens}, {Rogues}, {Rohrbasser}, {Romero-G{\'o}mez}, {Rowell}, {Royer}, {Ruz Mieres}, {Rybicki}, {Sadowski}, {S{\'a}ez N{\'u}{\~n}ez}, {Sagrist{\`a} Sell{\'e}s}, {Sahlmann}, {Salguero}, {Samaras}, {Sanchez Gimenez}, {Sanna}, {Santove{\~n}a}, {Sarasso}, {Schultheis}, {Sciacca}, {Segol}, {Segovia}, {S{\'e}gransan}, {Semeux}, {Shahaf}, {Siddiqui}, {Siebert},
  {Siltala}, {Silvelo}, {Slezak}, {Slezak}, {Smart}, {Snaith}, {Solano}, {Solitro}, {Souami}, {Souchay}, {Spagna}, {Spina}, {Spoto}, {Steele}, {Steidelm{\"u}ller}, {Stephenson}, {S{\"u}veges}, {Surdej}, {Szabados}, {Szegedi-Elek}, {Taris}, {Taylor}, {Teixeira}, {Tolomei}, {Tonello}, {Torra}, {Torra}, {Torralba Elipe}, {Trabucchi}, {Tsounis}, {Turon}, {Ulla}, {Unger}, {Vaillant}, {van Dillen}, {van Reeven}, {Vanel}, {Vecchiato}, {Viala}, {Vicente}, {Voutsinas}, {Weiler}, {Wevers}, {Wyrzykowski}, {Yoldas}, {Yvard}, {Zhao}, {Zorec}, {Zucker}, \& {Zwitter}}]{Gaia2023}
{Gaia Collaboration}, {Vallenari}, A., {Brown}, A.~G.~A., {et~al.} 2023, \aap, 674, A1

\bibitem[{{Gall} \& {Hjorth}(2018)}]{gall2018}
{Gall}, C. \& {Hjorth}, J. 2018, \apj, 868, 62

\bibitem[{{Gazagnes} {et~al.}(2018){Gazagnes}, {Chisholm}, {Schaerer}, {Verhamme}, {Rigby}, \& {Bayliss}}]{gazagnes2018}
{Gazagnes}, S., {Chisholm}, J., {Schaerer}, D., {et~al.} 2018, \aap, 616, A29

\bibitem[{{Gazagnes} {et~al.}(2024){Gazagnes}, {Cullen}, {Mauerhofer}, {Begley}, {Berg}, {Blaizot}, {Chisholm}, {Garel}, {Leclercq}, {McLure}, \& {Verhamme}}]{Gazagnes2024}
{Gazagnes}, S., {Cullen}, F., {Mauerhofer}, V., {et~al.} 2024, \apj, 969, 50

\bibitem[{{Glazebrook} {et~al.}(2024){Glazebrook}, {Nanayakkara}, {Schreiber}, {Lagos}, {Kawinwanichakij}, {Jacobs}, {Chittenden}, {Brammer}, {Kacprzak}, {Labbe}, {Marchesini}, {Marsan}, {Oesch}, {Papovich}, {Remus}, {Tran}, {Esdaile}, \& {Chandro-Gomez}}]{Glazebrook2024}
{Glazebrook}, K., {Nanayakkara}, T., {Schreiber}, C., {et~al.} 2024, \nat, 628, 277

\bibitem[{{Heckman} {et~al.}(2001){Heckman}, {Sembach}, {Meurer}, {Leitherer}, {Calzetti}, \& {Martin}}]{heckman2001}
{Heckman}, T.~M., {Sembach}, K.~R., {Meurer}, G.~R., {et~al.} 2001, \apj, 558, 56

\bibitem[{{Hegde} {et~al.}(2024){Hegde}, {Wyatt}, \& {Furlanetto}}]{Hegde2024}
{Hegde}, S., {Wyatt}, M.~M., \& {Furlanetto}, S.~R. 2024, arXiv e-prints, arXiv:2405.01629

\bibitem[{{Inayoshi} {et~al.}(2022){Inayoshi}, {Harikane}, {Inoue}, {Li}, \& {Ho}}]{Inayoshi2022}
{Inayoshi}, K., {Harikane}, Y., {Inoue}, A.~K., {Li}, W., \& {Ho}, L.~C. 2022, \apjl, 938, L10

\bibitem[{{Izotov} {et~al.}(2018){Izotov}, {Worseck}, {Schaerer}, {Guseva}, {Thuan}, {Fricke}, \& {Orlitov{\'a}}}]{izotov2018b}
{Izotov}, Y.~I., {Worseck}, G., {Schaerer}, D., {et~al.} 2018, \mnras, 478, 4851

\bibitem[{{Jaskot} \& {Oey}(2013)}]{jaskot2013}
{Jaskot}, A.~E. \& {Oey}, M.~S. 2013, \apj, 766, 91

\bibitem[{{Jaskot} {et~al.}(2017){Jaskot}, {Oey}, {Scarlata}, \& {Dowd}}]{Jaskot2017}
{Jaskot}, A.~E., {Oey}, M.~S., {Scarlata}, C., \& {Dowd}, T. 2017, \apjl, 851, L9

\bibitem[{{Jecmen} \& {Oey}(2023)}]{Jecmen2023}
{Jecmen}, M.~C. \& {Oey}, M.~S. 2023, \apj, 958, 149

\bibitem[{{Kannan} {et~al.}(2023){Kannan}, {Springel}, {Hernquist}, {Pakmor}, {Delgado}, {Hadzhiyska}, {Hern{\'a}ndez-Aguayo}, {Barrera}, {Ferlito}, {Bose}, {White}, {Frenk}, {Smith}, \& {Garaldi}}]{Kannan2023}
{Kannan}, R., {Springel}, V., {Hernquist}, L., {et~al.} 2023, \mnras, 524, 2594

\bibitem[{{Kimm} {et~al.}(2019){Kimm}, {Blaizot}, {Garel}, {Michel-Dansac}, {Katz}, {Rosdahl}, {Verhamme}, \& {Haehnelt}}]{Kimm2019}
{Kimm}, T., {Blaizot}, J., {Garel}, T., {et~al.} 2019, \mnras, 486, 2215

\bibitem[{{Komarova} {et~al.}(2021){Komarova}, {Oey}, {Krumholz}, {Silich}, {Kumari}, \& {James}}]{Komarova2021}
{Komarova}, L., {Oey}, M.~S., {Krumholz}, M.~R., {et~al.} 2021, \apjl, 920, L46

\bibitem[{{Krause} {et~al.}(2016){Krause}, {Charbonnel}, {Bastian}, \& {Diehl}}]{Krause2016}
{Krause}, M. G.~H., {Charbonnel}, C., {Bastian}, N., \& {Diehl}, R. 2016, \aap, 587, A53

\bibitem[{{Krause} {et~al.}(2020){Krause}, {Offner}, {Charbonnel}, {Gieles}, {Klessen}, {V{\'a}zquez-Semadeni}, {Ballesteros-Paredes}, {Girichidis}, {Kruijssen}, {Ward}, \& {Zinnecker}}]{Krause2020}
{Krause}, M. G.~H., {Offner}, S. S.~R., {Charbonnel}, C., {et~al.} 2020, \ssr, 216, 64

\bibitem[{{Kruijssen}(2012)}]{Kruijssen2012}
{Kruijssen}, J.~M.~D. 2012, \mnras, 426, 3008

\bibitem[{{Kusakabe} {et~al.}(2022){Kusakabe}, {Verhamme}, {Blaizot}, {Garel}, {Wisotzki}, {Leclercq}, {Bacon}, {Schaye}, {Gallego}, {Kerutt}, {Matthee}, {Maseda}, {Nanayakkara}, {Pell{\'o}}, {Richard}, {Tresse}, {Urrutia}, \& {Vitte}}]{Kusakabe2022}
{Kusakabe}, H., {Verhamme}, A., {Blaizot}, J., {et~al.} 2022, \aap, 660, A44

\bibitem[{{Langeroodi} \& {Hjorth}(2023)}]{Langeroodi2023}
{Langeroodi}, D. \& {Hjorth}, J. 2023, arXiv e-prints, arXiv:2307.06336

\bibitem[{{Larson}(1976)}]{Larson1976}
{Larson}, R.~B. 1976, \mnras, 176, 31

\bibitem[{{Leclercq} {et~al.}(2020){Leclercq}, {Bacon}, {Verhamme}, {Garel}, {Blaizot}, {Brinchmann}, {Cantalupo}, {Claeyssens}, {Conseil}, {Contini}, {Hashimoto}, {Herenz}, {Kusakabe}, {Marino}, {Maseda}, {Matthee}, {Mitchell}, {Pezzulli}, {Richard}, {Schmidt}, \& {Wisotzki}}]{Leclercq2020}
{Leclercq}, F., {Bacon}, R., {Verhamme}, A., {et~al.} 2020, \aap, 635, A82

\bibitem[{{Leclercq} {et~al.}(2017){Leclercq}, {Bacon}, {Wisotzki}, {Mitchell}, {Garel}, {Verhamme}, {Blaizot}, {Hashimoto}, {Herenz}, {Conseil}, {Cantalupo}, {Inami}, {Contini}, {Richard}, {Maseda}, {Schaye}, {Marino}, {Akhlaghi}, {Brinchmann}, \& {Carollo}}]{leclercq2017}
{Leclercq}, F., {Bacon}, R., {Wisotzki}, L., {et~al.} 2017, \aap, 608, A8

\bibitem[{{Li} {et~al.}(2023){Li}, {Dekel}, {Sarkar}, {Aung}, {Giavalisco}, {Mandelker}, \& {Tacchella}}]{Li2023}
{Li}, Z., {Dekel}, A., {Sarkar}, K.~C., {et~al.} 2023, arXiv e-prints, arXiv:2311.14662

\bibitem[{{Lotz} {et~al.}(2008){Lotz}, {Jonsson}, {Cox}, \& {Primack}}]{Lotz2008}
{Lotz}, J.~M., {Jonsson}, P., {Cox}, T.~J., \& {Primack}, J.~R. 2008, \mnras, 391, 1137

\bibitem[{{Lovell} {et~al.}(2023){Lovell}, {Harrison}, {Harikane}, {Tacchella}, \& {Wilkins}}]{Lovell2023MNRAS}
{Lovell}, C.~C., {Harrison}, I., {Harikane}, Y., {Tacchella}, S., \& {Wilkins}, S.~M. 2023, \mnras, 518, 2511

\bibitem[{{Magnelli} {et~al.}(2020){Magnelli}, {Boogaard}, {Decarli}, {G{\'o}nzalez-L{\'o}pez}, {Novak}, {Popping}, {Smail}, {Walter}, {Aravena}, {Assef}, {Bauer}, {Bertoldi}, {Carilli}, {Cortes}, {Cunha}, {Daddi}, {D{\'\i}az-Santos}, {Inami}, {Ivison}, {F{\`e}vre}, {Oesch}, {Riechers}, {Rix}, {Sargent}, {Werf}, {Wagg}, \& {Weiss}}]{Magnelli2020}
{Magnelli}, B., {Boogaard}, L., {Decarli}, R., {et~al.} 2020, \apj, 892, 66

\bibitem[{{Mainali} {et~al.}(2022){Mainali}, {Rigby}, {Chisholm}, {Bayliss}, {Bordoloi}, {Gladders}, {Rivera-Thorsen}, {Dahle}, {Sharon}, {Florian}, {Berg}, {Sharma}, {Owens}, {Kjellgren}, {Kim}, \& {Wayne}}]{Mainali2022}
{Mainali}, R., {Rigby}, J.~R., {Chisholm}, J., {et~al.} 2022, \apj, 940, 160

\bibitem[{{Maiolino} {et~al.}(2024){Maiolino}, {Scholtz}, {Witstok}, {Carniani}, {D'Eugenio}, {de Graaff}, {{\"U}bler}, {Tacchella}, {Curtis-Lake}, {Arribas}, {Bunker}, {Charlot}, {Chevallard}, {Curti}, {Looser}, {Maseda}, {Rawle}, {Rodr{\'\i}guez del Pino}, {Willott}, {Egami}, {Eisenstein}, {Hainline}, {Robertson}, {Williams}, {Willmer}, {Baker}, {Boyett}, {DeCoursey}, {Fabian}, {Helton}, {Ji}, {Jones}, {Kumari}, {Laporte}, {Nelson}, {Perna}, {Sandles}, {Shivaei}, \& {Sun}}]{Maiolino2024}
{Maiolino}, R., {Scholtz}, J., {Witstok}, J., {et~al.} 2024, \nat, 627, 59

\bibitem[{{Marques-Chaves} {et~al.}(2020{\natexlab{a}}){Marques-Chaves}, {{\'A}lvarez-M{\'a}rquez}, {Colina}, {P{\'e}rez-Fournon}, {Schaerer}, {Dalla Vecchia}, {Hashimoto}, {Jim{\'e}nez-{\'A}ngel}, \& {Shu}}]{marques2020b}
{Marques-Chaves}, R., {{\'A}lvarez-M{\'a}rquez}, J., {Colina}, L., {et~al.} 2020{\natexlab{a}}, \mnras, 499, L105

\bibitem[{{Marques-Chaves} {et~al.}(2020{\natexlab{b}}){Marques-Chaves}, {P{\'e}rez-Fournon}, {Shu}, {Colina}, {Bolton}, {{\'A}lvarez-M{\'a}rquez}, {Brownstein}, {Cornachione}, {Geier}, {Jim{\'e}nez-{\'A}ngel}, {Kojima}, {Mao}, {Montero-Dorta}, {Oguri}, {Ouchi}, {Poidevin}, {Shirley}, \& {Zheng}}]{marques2020}
{Marques-Chaves}, R., {P{\'e}rez-Fournon}, I., {Shu}, Y., {et~al.} 2020{\natexlab{b}}, \mnras, 492, 1257

\bibitem[{{Marques-Chaves} {et~al.}(2021){Marques-Chaves}, {Schaerer}, {{\'A}lvarez-M{\'a}rquez}, {Colina}, {Dessauges-Zavadsky}, {P{\'e}rez-Fournon}, {Saldana-Lopez}, \& {Verhamme}}]{marques2021}
{Marques-Chaves}, R., {Schaerer}, D., {{\'A}lvarez-M{\'a}rquez}, J., {et~al.} 2021, \mnras, 507, 524

\bibitem[{{Marques-Chaves} {et~al.}(2022){Marques-Chaves}, {Schaerer}, {{\'A}lvarez-M{\'a}rquez}, {Verhamme}, {Ceverino}, {Chisholm}, {Colina}, {Dessauges-Zavadsky}, {P{\'e}rez-Fournon}, {Saldana-Lopez}, {Upadhyaya}, \& {Vanzella}}]{marques2022}
{Marques-Chaves}, R., {Schaerer}, D., {{\'A}lvarez-M{\'a}rquez}, J., {et~al.} 2022, \mnras, 517, 2972

\bibitem[{{Marques-Chaves} {et~al.}(2024){Marques-Chaves}, {Schaerer}, {Kuruvanthodi}, {Korber}, {Prantzos}, {Charbonnel}, {Weibel}, {Izotov}, {Messa}, {Brammer}, {Dessauges-Zavadsky}, \& {Oesch}}]{Marques2024}
{Marques-Chaves}, R., {Schaerer}, D., {Kuruvanthodi}, A., {et~al.} 2024, \aap, 681, A30

\bibitem[{{Martin} {et~al.}(2015){Martin}, {Dijkstra}, {Henry}, {Soto}, {Danforth}, \& {Wong}}]{martin2015}
{Martin}, C.~L., {Dijkstra}, M., {Henry}, A., {et~al.} 2015, \apj, 803, 6

\bibitem[{{Martins} \& {Palacios}(2022)}]{Martins2022}
{Martins}, F. \& {Palacios}, A. 2022, \aap, 659, A163

\bibitem[{{Martins} {et~al.}(2023){Martins}, {Schaerer}, {Marques-Chaves}, \& {Upadhyaya}}]{Martins2023}
{Martins}, F., {Schaerer}, D., {Marques-Chaves}, R., \& {Upadhyaya}, A. 2023, \aap, 678, A159

\bibitem[{{Mason} {et~al.}(2023){Mason}, {Trenti}, \& {Treu}}]{Mason2023}
{Mason}, C.~A., {Trenti}, M., \& {Treu}, T. 2023, \mnras, 521, 497

\bibitem[{{Matteucci}(1994)}]{Matteucci1994}
{Matteucci}, F. 1994, \aap, 288, 57

\bibitem[{{Mauerhofer} {et~al.}(2021){Mauerhofer}, {Verhamme}, {Blaizot}, {Garel}, {Kimm}, {Michel-Dansac}, \& {Rosdahl}}]{mauerhofer2021}
{Mauerhofer}, V., {Verhamme}, A., {Blaizot}, J., {et~al.} 2021, \aap, 646, A80

\bibitem[{{Megeath} {et~al.}(2016){Megeath}, {Gutermuth}, {Muzerolle}, {Kryukova}, {Hora}, {Allen}, {Flaherty}, {Hartmann}, {Myers}, {Pipher}, {Stauffer}, {Young}, \& {Fazio}}]{Megeath2016}
{Megeath}, S.~T., {Gutermuth}, R., {Muzerolle}, J., {et~al.} 2016, \aj, 151, 5

\bibitem[{{Menon} {et~al.}(2024{\natexlab{a}}){Menon}, {Burkhart}, {Somerville}, {Thompson}, \& {Sternberg}}]{Menon2024b}
{Menon}, S.~H., {Burkhart}, B., {Somerville}, R.~S., {Thompson}, T.~A., \& {Sternberg}, A. 2024{\natexlab{a}}, arXiv e-prints, arXiv:2408.14591

\bibitem[{{Menon} {et~al.}(2023){Menon}, {Federrath}, \& {Krumholz}}]{Menon2023}
{Menon}, S.~H., {Federrath}, C., \& {Krumholz}, M.~R. 2023, \mnras, 521, 5160

\bibitem[{{Menon} {et~al.}(2024{\natexlab{b}}){Menon}, {Lancaster}, {Burkhart}, {Somerville}, {Dekel}, \& {Krumholz}}]{Menon2024}
{Menon}, S.~H., {Lancaster}, L., {Burkhart}, B., {et~al.} 2024{\natexlab{b}}, \apjl, 967, L28

\bibitem[{{Messa} {et~al.}(2024){Messa}, {Dessauges-Zavadsky}, {Adamo}, {Richard}, \& {Claeyssens}}]{Messa2024}
{Messa}, M., {Dessauges-Zavadsky}, M., {Adamo}, A., {Richard}, J., \& {Claeyssens}, A. 2024, \mnras, 529, 2162

\bibitem[{{Messa} {et~al.}(2022){Messa}, {Dessauges-Zavadsky}, {Richard}, {Adamo}, {Nagy}, {Combes}, {Mayer}, \& {Ebeling}}]{Messa2022}
{Messa}, M., {Dessauges-Zavadsky}, M., {Richard}, J., {et~al.} 2022, \mnras, 516, 2420

\bibitem[{{Me{\v{s}}tri{\'c}} {et~al.}(2023){Me{\v{s}}tri{\'c}}, {Vanzella}, {Upadhyaya}, {Martins}, {Marques-Chaves}, {Schaerer}, {Guibert}, {Zanella}, {Grillo}, {Rosati}, {Calura}, {Caminha}, {Bolamperti}, {Meneghetti}, {Bergamini}, {Mercurio}, {Nonino}, \& {Pascale}}]{Mestric2023}
{Me{\v{s}}tri{\'c}}, U., {Vanzella}, E., {Upadhyaya}, A., {et~al.} 2023, \aap, 673, A50

\bibitem[{{Morishita} {et~al.}(2024){Morishita}, {Stiavelli}, {Chary}, {Trenti}, {Bergamini}, {Chiaberge}, {Leethochawalit}, {Roberts-Borsani}, {Shen}, \& {Treu}}]{Morishita2024}
{Morishita}, T., {Stiavelli}, M., {Chary}, R.-R., {et~al.} 2024, \apj, 963, 9

\bibitem[{{Norris} {et~al.}(2014){Norris}, {Kannappan}, {Forbes}, {Romanowsky}, {Brodie}, {Faifer}, {Huxor}, {Maraston}, {Moffett}, {Penny}, {Pota}, {Smith-Castelli}, {Strader}, {Bradley}, {Eckert}, {Fohring}, {McBride}, {Stark}, \& {Vaduvescu}}]{norris2014}
{Norris}, M.~A., {Kannappan}, S.~J., {Forbes}, D.~A., {et~al.} 2014, \mnras, 443, 1151

\bibitem[{{Oteo} {et~al.}(2017){Oteo}, {Zwaan}, {Ivison}, {Smail}, \& {Biggs}}]{Oteo2017}
{Oteo}, I., {Zwaan}, M.~A., {Ivison}, R.~J., {Smail}, I., \& {Biggs}, A.~D. 2017, \apj, 837, 182

\bibitem[{{Pasha} \& {Miller}(2023)}]{Pasha2023}
{Pasha}, I. \& {Miller}, T.~B. 2023, The Journal of Open Source Software, 8, 5703

\bibitem[{{Renzini}(2023)}]{Renzini2023}
{Renzini}, A. 2023, \mnras, 525, L117

\bibitem[{{Ribeiro} {et~al.}(2016){Ribeiro}, {Le F{\`e}vre}, {Tasca}, {Lemaux}, {Cassata}, {Garilli}, {Maccagni}, {Zamorani}, {Zucca}, {Amor{\'{\i}}n}, {Bardelli}, {Fontana}, {Giavalisco}, {Hathi}, {Koekemoer}, {Pforr}, {Tresse}, \& {Dunlop}}]{ribeiro2016}
{Ribeiro}, B., {Le F{\`e}vre}, O., {Tasca}, L.~A.~M., {et~al.} 2016, \aap, 593, A22

\bibitem[{{Rivera-Thorsen} {et~al.}(2019){Rivera-Thorsen}, {Dahle}, {Chisholm}, {Florian}, {Gronke}, {Rigby}, {Gladders}, {Mahler}, {Sharon}, \& {Bayliss}}]{rivera2019}
{Rivera-Thorsen}, T.~E., {Dahle}, H., {Chisholm}, J., {et~al.} 2019, Science, 366, 738

\bibitem[{{Rivera-Thorsen} {et~al.}(2017){Rivera-Thorsen}, {Dahle}, {Gronke}, {Bayliss}, {Rigby}, {Simcoe}, {Bordoloi}, {Turner}, \& {Furesz}}]{rivera2017}
{Rivera-Thorsen}, T.~E., {Dahle}, H., {Gronke}, M., {et~al.} 2017, \aap, 608, L4

\bibitem[{{Saldana-Lopez} {et~al.}(2022){Saldana-Lopez}, {Schaerer}, {Chisholm}, {Flury}, {Jaskot}, {Worseck}, {Makan}, {Gazagnes}, {Mauerhofer}, {Verhamme}, {Amor{\'\i}n}, {Ferguson}, {Giavalisco}, {Grazian}, {Hayes}, {Heckman}, {Henry}, {Ji}, {Marques-Chaves}, {McCandliss}, {Oey}, {{\"O}stlin}, {Pentericci}, {Thuan}, {Trebitsch}, {Vanzella}, \& {Xu}}]{saldana2022}
{Saldana-Lopez}, A., {Schaerer}, D., {Chisholm}, J., {et~al.} 2022, \aap, 663, A59

\bibitem[{{Schaerer} {et~al.}(2024{\natexlab{a}}){Schaerer}, {Guibert}, {Marques-Chaves}, \& {Martins}}]{Schaerer2024b}
{Schaerer}, D., {Guibert}, J., {Marques-Chaves}, R., \& {Martins}, F. 2024{\natexlab{a}}, arXiv e-prints, arXiv:2407.12122

\bibitem[{{Schaerer} {et~al.}(2024{\natexlab{b}}){Schaerer}, {Marques-Chaves}, {Xiao}, \& {Korber}}]{Schaerer2024}
{Schaerer}, D., {Marques-Chaves}, R., {Xiao}, M., \& {Korber}, D. 2024{\natexlab{b}}, \aap, 687, L11

\bibitem[{{Senchyna} {et~al.}(2024){Senchyna}, {Plat}, {Stark}, {Rudie}, {Berg}, {Charlot}, {James}, \& {Mingozzi}}]{Senchyna2024}
{Senchyna}, P., {Plat}, A., {Stark}, D.~P., {et~al.} 2024, \apj, 966, 92

\bibitem[{{Shapley} {et~al.}(2003){Shapley}, {Steidel}, {Pettini}, \& {Adelberger}}]{shapley2003}
{Shapley}, A.~E., {Steidel}, C.~C., {Pettini}, M., \& {Adelberger}, K.~L. 2003, \apj, 588, 65

\bibitem[{{Shen} {et~al.}(2023){Shen}, {Vogelsberger}, {Boylan-Kolchin}, {Tacchella}, \& {Kannan}}]{Shen2023}
{Shen}, X., {Vogelsberger}, M., {Boylan-Kolchin}, M., {Tacchella}, S., \& {Kannan}, R. 2023, \mnras, 525, 3254

\bibitem[{{Smith} {et~al.}(2018){Smith}, {Windhorst}, {Jansen}, {Cohen}, {Jiang}, {Dijkstra}, {Koekemoer}, {Bielby}, {Inoue}, {MacKenty}, {O'Connell}, \& {Silk}}]{Smith2018}
{Smith}, B.~M., {Windhorst}, R.~A., {Jansen}, R.~A., {et~al.} 2018, \apj, 853, 191

\bibitem[{{Stanway} \& {Eldridge}(2018)}]{stanway2018}
{Stanway}, E.~R. \& {Eldridge}, J.~J. 2018, \mnras, 479, 75

\bibitem[{{Steidel} {et~al.}(2018){Steidel}, {Bogosavljevi{\'c}}, {Shapley}, {Reddy}, {Rudie}, {Pettini}, {Trainor}, \& {Strom}}]{steidel2018}
{Steidel}, C.~C., {Bogosavljevi{\'c}}, M., {Shapley}, A.~E., {et~al.} 2018, \apj, 869, 123

\bibitem[{{Steidel} {et~al.}(2010){Steidel}, {Erb}, {Shapley}, {Pettini}, {Reddy}, {Bogosavljevi{\'c}}, {Rudie}, \& {Rakic}}]{steidel2010}
{Steidel}, C.~C., {Erb}, D.~K., {Shapley}, A.~E., {et~al.} 2010, \apj, 717, 289

\bibitem[{{Tacchella} {et~al.}(2016){Tacchella}, {Dekel}, {Carollo}, {Ceverino}, {DeGraf}, {Lapiner}, {Mandelker}, \& {Primack}}]{tacchella2016}
{Tacchella}, S., {Dekel}, A., {Carollo}, C.~M., {et~al.} 2016, \mnras, 458, 242

\bibitem[{{Tang} {et~al.}(2022){Tang}, {Stark}, \& {Ellis}}]{Tang2022}
{Tang}, M., {Stark}, D.~P., \& {Ellis}, R.~S. 2022, \mnras, 513, 5211

\bibitem[{{Thompson} \& {Heckman}(2024)}]{Thompson2024}
{Thompson}, T.~A. \& {Heckman}, T.~M. 2024, arXiv e-prints, arXiv:2406.08561

\bibitem[{{Topping} {et~al.}(2024){Topping}, {Stark}, {Senchyna}, {Plat}, {Zitrin}, {Endsley}, {Charlot}, {Furtak}, {Maseda}, {Smit}, {Mainali}, {Chevallard}, {Molyneux}, \& {Rigby}}]{Topping2024}
{Topping}, M.~W., {Stark}, D.~P., {Senchyna}, P., {et~al.} 2024, \mnras, 529, 3301

\bibitem[{{Trinca} {et~al.}(2024){Trinca}, {Schneider}, {Valiante}, {Graziani}, {Ferrotti}, {Omukai}, \& {Chon}}]{Trinca2024}
{Trinca}, A., {Schneider}, R., {Valiante}, R., {et~al.} 2024, \mnras, 529, 3563

\bibitem[{{Upadhyaya} {et~al.}(2024){Upadhyaya}, {Marques-Chaves}, {Schaerer}, {Martins}, {P{\'e}rez-Fournon}, {Palacios}, \& {Stanway}}]{Upadhyaya2024}
{Upadhyaya}, A., {Marques-Chaves}, R., {Schaerer}, D., {et~al.} 2024, \aap, 686, A185

\bibitem[{{van der Wel} {et~al.}(2012){van der Wel}, {Bell}, {H{\"a}ussler}, {McGrath}, {Chang}, {Guo}, {McIntosh}, {Rix}, {Barden}, {Cheung}, {Faber}, {Ferguson}, {Galametz}, {Grogin}, {Hartley}, {Kartaltepe}, {Kocevski}, {Koekemoer}, {Lotz}, {Mozena}, {Peth}, \& {Peng}}]{van2012}
{van der Wel}, A., {Bell}, E.~F., {H{\"a}ussler}, B., {et~al.} 2012, \apjs, 203, 24

\bibitem[{{van Dokkum} {et~al.}(2008){van Dokkum}, {Franx}, {Kriek}, {Holden}, {Illingworth}, {Magee}, {Bouwens}, {Marchesini}, {Quadri}, {Rudnick}, {Taylor}, \& {Toft}}]{vandokkum2008}
{van Dokkum}, P.~G., {Franx}, M., {Kriek}, M., {et~al.} 2008, \apjl, 677, L5

\bibitem[{{Vanzella} {et~al.}(2020){Vanzella}, {Caminha}, {Calura}, {Cupani}, {Meneghetti}, {Castellano}, {Rosati}, {Mercurio}, {Sani}, {Grillo}, {Gilli}, {Mignoli}, {Comastri}, {Nonino}, {Cristiani}, {Giavalisco}, \& {Caputi}}]{vanzella2020}
{Vanzella}, E., {Caminha}, G.~B., {Calura}, F., {et~al.} 2020, \mnras, 491, 1093

\bibitem[{{Vanzella} {et~al.}(2022){Vanzella}, {Castellano}, {Bergamini}, {Meneghetti}, {Zanella}, {Calura}, {Caminha}, {Rosati}, {Cupani}, {Me{\v{s}}tri{\'c}}, {Brammer}, {Tozzi}, {Mercurio}, {Grillo}, {Sani}, {Cristiani}, {Nonino}, {Merlin}, \& {Pignataro}}]{Vanzella2022}
{Vanzella}, E., {Castellano}, M., {Bergamini}, P., {et~al.} 2022, \aap, 659, A2

\bibitem[{{Vanzella} {et~al.}(2023){Vanzella}, {Claeyssens}, {Welch}, {Adamo}, {Coe}, {Diego}, {Mahler}, {Khullar}, {Kokorev}, {Oguri}, {Ravindranath}, {Furtak}, {Hsiao}, {Abdurro'uf}, {Mandelker}, {Brammer}, {Bradley}, {Brada{\v{c}}}, {Conselice}, {Dayal}, {Nonino}, {Andrade-Santos}, {Windhorst}, {Pirzkal}, {Sharon}, {de Mink}, {Fujimoto}, {Zitrin}, {Eldridge}, \& {Norman}}]{vanzella2023}
{Vanzella}, E., {Claeyssens}, A., {Welch}, B., {et~al.} 2023, \apj, 945, 53

\bibitem[{{Verhamme} {et~al.}(2012){Verhamme}, {Dubois}, {Blaizot}, {Garel}, {Bacon}, {Devriendt}, {Guiderdoni}, \& {Slyz}}]{Verhamme2012}
{Verhamme}, A., {Dubois}, Y., {Blaizot}, J., {et~al.} 2012, \aap, 546, A111

\bibitem[{{Verhamme} {et~al.}(2015){Verhamme}, {Orlitov{\'a}}, {Schaerer}, \& {Hayes}}]{verhamme2015}
{Verhamme}, A., {Orlitov{\'a}}, I., {Schaerer}, D., \& {Hayes}, M. 2015, Astronomy and Astrophysics, 578, A7

\bibitem[{{Weibel} {et~al.}(2024){Weibel}, {Oesch}, {Barrufet}, {Gottumukkala}, {Ellis}, {Santini}, {Weaver}, {Allen}, {Bouwens}, {Bowler}, {Brammer}, {Carnall}, {Cullen}, {Dayal}, {Donnan}, {Dunlop}, {Giavalisco}, {Grogin}, {Illingworth}, {Koekemoer}, {Labbe}, {Marchesini}, {McLeod}, {McLure}, {Naidu}, {Shuntov}, {Stefanon}, {Toft}, \& {Xiao}}]{Weibel2024}
{Weibel}, A., {Oesch}, P.~A., {Barrufet}, L., {et~al.} 2024, arXiv e-prints, arXiv:2403.08872

\bibitem[{{Williams} {et~al.}(2023){Williams}, {Kelly}, {Chen}, {Brammer}, {Zitrin}, {Treu}, {Scarlata}, {Koekemoer}, {Oguri}, {Lin}, {Diego}, {Nonino}, {Hjorth}, {Langeroodi}, {Broadhurst}, {Rogers}, {Perez-Fournon}, {Foley}, {Jha}, {Filippenko}, {Strolger}, {Pierel}, {Poidevin}, \& {Yang}}]{Williams2023}
{Williams}, H., {Kelly}, P.~L., {Chen}, W., {et~al.} 2023, Science, 380, 416

\bibitem[{{Xiao} {et~al.}(2023){Xiao}, {Oesch}, {Elbaz}, {Bing}, {Nelson}, {Weibel}, {Naidu}, {Daddi}, {Bouwens}, {Matthee}, {Wuyts}, {Chisholm}, {Brammer}, {Dickinson}, {Magnelli}, {Leroy}, {van Dokkum}, {Schaerer}, {Herard-Demanche}, {Barrufet}, {Endsley}, {Fudamoto}, {G{\'o}mez-Guijarro}, {Gottumukkala}, {Illingworth}, {Labbe}, {Magee}, {Marchesini}, {Maseda}, {Qin}, {Reddy}, {Shapley}, {Shivaei}, {Shuntov}, {Stefanon}, {Whitaker}, \& {Wyithe}}]{Xiao2023}
{Xiao}, M., {Oesch}, P., {Elbaz}, D., {et~al.} 2023, arXiv e-prints, arXiv:2309.02492

\bibitem[{{Ziparo} {et~al.}(2023){Ziparo}, {Ferrara}, {Sommovigo}, \& {Kohandel}}]{Ziparo2023}
{Ziparo}, F., {Ferrara}, A., {Sommovigo}, L., \& {Kohandel}, M. 2023, \mnras, 520, 2445

\bibitem[{{Zolotov} {et~al.}(2015){Zolotov}, {Dekel}, {Mandelker}, {Tweed}, {Inoue}, {DeGraf}, {Ceverino}, {Primack}, {Barro}, \& {Faber}}]{zolotov2015}
{Zolotov}, A., {Dekel}, A., {Mandelker}, N., {et~al.} 2015, \mnras, 450, 2327

\end{thebibliography}

\end{document}